\documentclass[oldversion]{aa} 
\usepackage{graphicx}
\usepackage{aalongtable}
\usepackage{txfonts}
\usepackage{rotating}
\usepackage{lscape}
\newcommand{\gsim}{\;\lower.6ex\hbox{$\sim$}\kern-7.75pt\raise.65ex\hbox{$>$}\;}
\newcommand{\lsim}{\;\lower.6ex\hbox{$\sim$}\kern-7.75pt\raise.65ex\hbox{$<$}\;}

\begin{document}
\title{Observing multiple populations in globular clusters with the ESO archive:
NGC~6388 reloaded\thanks{Based on observations collected at 
ESO telescopes under programmes 073.D-0211 (proprietary), and 073.D-0760,  381.D-0329,  095.D-0834 (archival)}
 }

\author{
Eugenio Carretta\inst{1}
\and
Angela Bragaglia\inst{1}
}

\authorrunning{Carretta and Bragaglia}
\titlerunning{Multiple populations in NGC~6388}

\offprints{E. Carretta, eugenio.carretta@oabo.inaf.it}

\institute{
INAF-Osservatorio di Astrofisica e Scienza dello Spazio di Bologna, Via Gobetti
 93/3, I-40129 Bologna, Italy}

\date{}

\abstract{The metal-rich and old bulge globular cluster (GC) NGC 6388 is one of
the most massive Galactic GCs ($M \sim 10^6 M_\odot$). However, the
spectroscopic properties of its multiple stellar populations  rested only on 32
red giants (only seven of which observed with UVES, the remaining with GIRAFFE),
given the difficulties in observing a rather distant cluster,  heavily
contaminated by bulge and disc field stars. We bypassed the problem using the
largest telescope facility ever: the European Southern Observatory (ESO)
archive. By selecting member stars identified by other programmes, we derive
atmospheric parameters and the full set of abundances for 15 species from high
resolution UVES spectra of  another 17 red giant branch stars in NGC~6388. We
confirm that no metallicity dispersion is appreciable in this GC. About 30\% of
stars show the primordial composition of first generation stars, about 20\%
present an extremely modified second generation composition, and half of the
stars has an  intermediate composition. The stars clearly distribute in the Al-O
and Na-O  planes into three discrete groups. We find substantial hints that more
than a single class of polluters is required to reproduce the composition of the
intermediate component in NGC~6388. In the heavily polluted component the sum
Mg+Al increases as Al increases. The sum Mg+Al+Si is constant, and  is the
fossil record of hot H-burning at temperatures higher than about 70 MK in the
first generation polluters that contributed to form multiple populations in this
cluster.
}

\maketitle

\section{Introduction}
Almost all Galactic globular clusters (GCs) are composed by multiple stellar
populations (see e.g. Gratton, Carretta, Bragaglia 2012 and Bastian and Lardo
2017 for overviews, and Bragaglia et al. 2017 for an updated census). This
represents both a challenge to any model of GC formation, and a strong
opportunity to understand the early evolution of GCs. The general outline of the
first phases of the GCs lifetime seems currently to be clear (although
there are some differing views, e.g. the presence of a single generation, 
Bastian et al. 2013): a first
stellar  generation is formed by gas only having the composition imprinted by
type II supernovae nucleosynthesis (overabundance of $\alpha$ elements,
including oxygen, underabundance of other light elements such as Na, Al). The
most massive stars in this first burst of star formation then evolve and
contribute nuclearly processed matter (rich in elements from proton-capture
reactions) to the gas pool from which a second stellar generation is formed with
a mix of enriched  ejecta and pristine gas. The ubiquitous anticorrelations C-N,
Na-O, Mg-Al found in GCs (e.g. Smith 1987, Carretta et al. 2009a,b, M\'esz\'aros
et al. 2015) leave no doubts that this is the basic sequence of events at the GC
birth time. What we still don't know is the exact nature of the first generation
polluters (e.g., intermediate-mass asymptotic giant branch AGB stars: Cottrell
\& Da Costa 1981; Ventura et al. 2001; fast rotating massive stars: Decressin et
al. 2007; massive  binaries: de Mink et al. 2009), and the sequence and duration
of the feedback-regulated secondary star formation.

In this framework, each GC seems to have a different star formation history and
the metal rich ([Fe/H]=-0.44, Carretta et al. 2007a), massive ($M_V=-9.41$ mag,
Harris 1996, 2010 edition) bulge cluster NGC~6388 certainly deserves attention
due to its peculiar features: (a) beside the red horizontal branch (HB) typical
of old, metal-rich GCs, NGC 6388 shows an extended blue HB, i.e., we see the
second parameter at work within the same cluster. The He enrichment in the
second generation stars, as suggested by D'Antona \& Caloi (2004), is likely 
the explanation, as shown by NGC~6388 participating in the
strong correlation between extension of the Na-O anticorrelation and maximum
temperature on the HB (discovered by Carretta et al. 2007b); (b) NGC 6388 is a
local counterpart of old, metal-rich populations found in distant elliptical
galaxies, and its relevant population of hot HB stars is a likely contributor to
the UV-upturn phenomenon (e.g. Yi, Demarque, Oemler 1998); (c) NGC 6388 is a
pivotal cluster to confirm the existence of intermediate-mass black holes (IMBH
$10^3-10^4 M_\odot$) whose evidence in this GC is controversial (Lanzoni et al.
2013, hereinafter L13;  Lutzgendorf et al. 2015).

\begin{figure}
\centering
\includegraphics[bb=30 150 400 700,clip,scale=0.40]{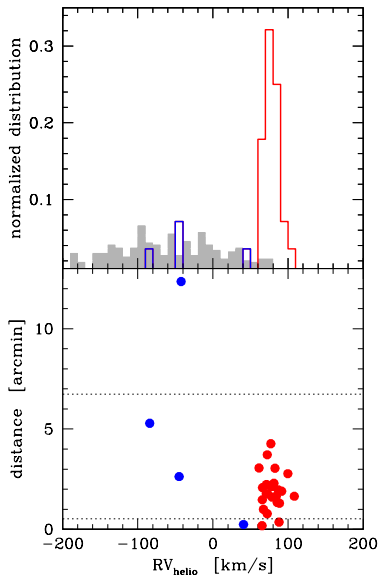}
\caption{Upper panel - Distribution of heliocentric RVs for our sample (open
histogram) and the tile closest to the position of NGC~6388 in Zoccali et al.
(2017, gray histogram), both normalized to the total numbers in each sample (28
and 437 stars, respectively). Bottom panel - RVs in NGC~6388 as a function of
the distance from the cluster centre (red: stars classified member, blue: non
member). The half-mass and tidal radii are indicated by the horizontal lines.}
\label{f:RV}
\end{figure}

The main observational problem in studying the composite stellar populations in
NGC~6388 comes from the strong contamination by disc and bulge stars. In our
FLAMES survey to study the Na-O anticorrelation in a large sample of GCs (see
Carretta et al. 2006) we were the first to obtain high-resolution spectra of 
many stars in NGC~6388, using FLAMES (Carretta et al. 2007a,2009b), but it was 
a shot in the dark as far as members were concerned and about one half of
the stars had to be discarded.
We could measure O, Na only in 32 member stars along the red giant branch (RGB),
seven of which with UVES, almost the lowest number in our FLAMES GC survey. The other two GCs with comparably low numbers of stars
measured with respect to their total mass were the bulge GC NGC~6441 (a twin of
NGC~6388 also in this regard) and the very metal-poor M~15 (NGC~7078).

Despite this limitation, we were able to find a conspicuous fraction (19\%) of
second generation stars with extremely modified composition, rich in Na (and
likely He), poor in O  in NGC~6388. These extreme stars are not found in all
GCs; their fraction is second only to M~54 and similar to the very peculiar GC
NGC~2808 (Carretta 2015). Clearly, this should be corroborated by more robust
statistics. Larger samples would also be useful to establish beyond any doubt
that NGC~6388 does not belong to the increasing class of the so-called {\it iron
complex} GCs (see Johnson et al. 2015, Marino et al. 2015 and references
therein), characterized by even consistent metallicity dispersion and correlated
[Fe/H]\footnote{We adopt the usual spectroscopic notation, $i.e.$  [X]=
log(X)$_{\rm star} -$ log(X)$_\odot$ for any abundance quantity X, and  log
$\epsilon$(X) = log (N$_{\rm X}$/N$_{\rm H}$) + 12.0 for absolute number density
abundances.} and $s-$process elements enhancements. All GCs in this class are
among the most massive in the Milky Way, hence NGC~6388 classifies as a good
candidate, even if no such signature was found in our first analysis (Carretta
et al. 2007a, 2009b). 

To improve the pool of member stars in NGC~6388 we took advantage of the huge 
potential of the ESO archive, where more than 300 stars, observed mostly with
FLAMES high-resolution GIRAFFE setup HR21, are available for this cluster. While
these data were originally acquired to explore the velocity dispersion (L13),
part of the spectra may be also exploited to derive abundances. In the present
work we started by analysing the high resolution UVES spectra taken within
different programmes for 17 new RGB stars in NGC~6388. The  large spectral
coverage of UVES ensures a large number of transitions for a variety of
species.  Hence we can provide abundances of elements from proton-captures to
improve the chemical characterization of multiple populations in NGC~6388, and
possibly to investigate the existence of discrete components. The [Fe/H]
distribution of the enlarged sample may confirm (or falsify) the conclusion that
no intrinsic metallicity dispersion is present in this cluster. The elements
from $\alpha-$captures and those belonging to the Fe-peak may be compared to
those of disc and bulge stars.

We organized the paper as follows: the adopted datasets from the ESO archive are
described in Sect. 2, whereas the analysis and the error budget are discussed in
Sect. 3 and 4, respectively. Sect. 5 is devoted to the results on multiple
populations in NGC~6388 and Sect. 6 to a brief description of the other
elements derived in the analysis. Finally, in Sect. 7 we summarize our findings
and illustrate future prospects.

\section{Adopted datasets}

Our present sample was assembled from our private data (Carretta et al. 2007a,
2009b) and the ESO archival spectra of three programmes: 381.D-0329(B), PI
Lanzoni;  073.D-0760(A), PI Catelan; 095.D-0834(A), PI Henault-Brunet;
hereinafter samples L13, CAT, and H-B, respectively. The first was only devoted
to search for kinematical signatures of IMBH in NGC~6388, and results were
published in L13. The two other programmes  were both aimed also to explore the
chemistry of this GC. 

We consider here only the UVES spectra. UVES observation from L13 include four
fibre configurations, one of them repeated twice, for a total of eight stars
(one is a non-member from its radial velocity, RV -see below). The H-B  UVES
sample includes six stars, repeated ten times, two of which are non-members.
Finally, the CAT data consists of three exposures, but only two are usable, with
two fibre configurations and a total of 14 observed stars, including one
non-member. 

We considered as good candidate member stars of NGC~6388 all the objects
with heliocentric RV between 60 and 110 km s$^{-1}$. Beside being in agreement 
with the limits adopted by the extensive work by L13 (their figure 4 and section
3.1), this interval shows a clear peak in the RV distribution
(Fig.~\ref{f:RV}, upper panel). As a comparison, we show the distribution of
bulge stars from Zoccali et al. (2017),  selecting only the stars closer in
latitude and longitude to NGC~6388. The two normalized distributions indicate
that our choice of candidate member stars is robust. Twenty four out of 28 stars
are probable members and we show their RV as function of the distance from the
cluster centre in the bottom panel of Fig.~\ref{f:RV}. One of the non members is
outside the cluster tidal radius, while the majority of observed stars lie
between the half-mass and tidal radii (both values from Harris 1996). Larger
samples, especially close to the centre, such as in L13, would describe better
the radial properties.

Neglecting stars with spectra having a S/N too low for a reliable abundance 
analysis (S/N$\sim10$ in the combined spectrum), we finally considered 11 stars
from CAT, five from L13 and only one from H-B, for a total of 17 RGB stars.
The final S/N per pixel are between $\sim 20$ and $\sim 50$, enough for
relatively cool stars in a metal-rich GC.

Relevant information (star ID, sample, original star names, coordinates,
magnitudes and heliocentric RVs) are given in Table~\ref{t:sample}. Since
different naming convention are used in the original samples, we adopted here a
new homogeneous convention, for sake of simplicity, which is Ann, A for archive
and nn a number from 1 to 17. Table~\ref{t:sample} lists the correspondence
with the original star names as well as with those of the photometry source.

\begin{table*}
\centering
\caption[]{Information on stars analyzed in NGC~6388}
\begin{tabular}{llrrccllrrr}
\hline
ID     & Sample & name   & name &RA  & DEC & $~~~V$  & $~~~B$ & $K~~~~~$ & RV Hel. &errRV    \\
(us)   &	& (orig) &(WFI) &    &	   & (mag)&(mag)&(mag)& km s$^{-1}$ & km s$^{-1}$\\
\hline        
\hline
A01& CAT &    1043 & 104258 & 264.119676 &-44.724892 &  15.237 &17.246 &10.081 & 79.46  &0.49\\
A02& CAT &    1063 & 102175 & 264.109499 &-44.736583 &  14.962 &16.979 & 9.824 & 78.83  &0.15\\
A03& CAT &    1165 &  98155 & 264.068602 &-44.759965 &  15.343 &16.993 &11.238 & 65.72  &0.25\\
A04& CAT &    1253 & 102159 & 264.033217 &-44.736818 &  14.757 &16.678 & 9.731 & 84.30  &0.93\\
A05& CAT &    1317 & 107018 & 264.050041 &-44.706611 &  15.337 &17.249 &10.493 & 86.93  &0.45\\
A06& CAT &    1337 & 107103 & 264.069666 &-44.705944 &  15.161 &17.095 &10.218 & 71.08  &0.10\\
A07& CAT &    2085 &  95520 & 264.107749 &-44.779694 &  15.052 &17.059 & 9.979 & 61.39  &0.43\\
A08& CAT &    1327 & 107702 & 264.050124 &-44.701499 &  15.198 &17.19  &10.125 & 71.42  &0.21\\
A09& CAT &    1153 &  96816 & 264.085124 &-44.768861 &  15.249 &16.986 &10.978 & 66.10  &0.14\\
A10& CAT &    1144 & 	    & 264.088583 &-44.762305 &  17.311 &       &10.477 & 72.79  &0.10\\
A11& CAT &    1189 &  97133 & 264.047416 &-44.766722 &  15.233 &17.033 &10.707 & 77.40  &0.40\\
A12& H-B &  181624 & 109894 & 264.009291 &-44.679916 &  16.493 &18.108 &12.353 & 77.12  &0.16\\
A13& L13 &  200316 & 103385 & 264.033953 &-44.729977 &  16.245 &17.832 &11.620 &108.37  &0.65\\
A14& L13 &  200340 & 	    & 264.054198 &-44.739107 &  16.30  &       &12.32  & 72.23  &0.53\\
A15& L13 &  200395 &  99213 & 264.054667 &-44.753447 &  16.359 &17.832 &12.611 & 88.59  &0.73\\
A16& L13 &   22133 & 104922 & 264.082833 &-44.720916 &  16.349 &17.814 &12.137 & 67.25  &0.20\\
A17& L13 & 7001303 & 104554 & 264.112892 &-44.723129 &  16.298 &17.836 &12.235 & 91.56  &0.40\\
\hline
\hline
\end{tabular}
\label{t:sample}
\end{table*}

\begin{figure}
\centering
\includegraphics[scale=0.40]{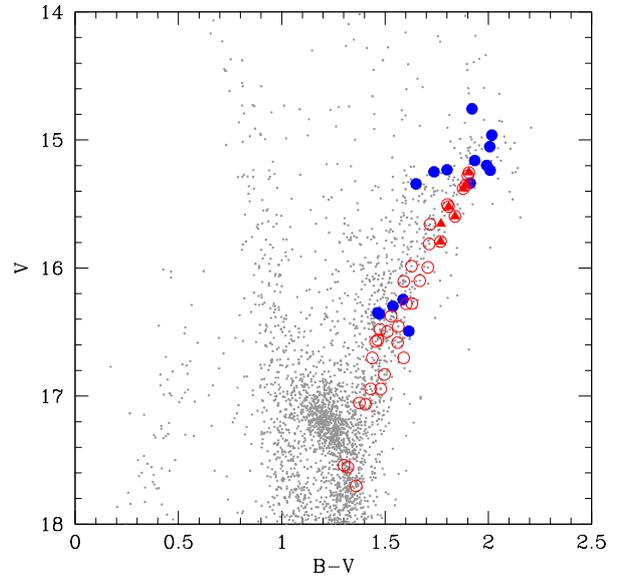}
\caption{$V,B-V$ colour-magnitude diagram for NGC~6388 using WFI photometry
(small grey points). Large blue filled circle indicate our new sample with UVES
spectra. Filled red triangles are giants with UVES spectra analyzed in Carretta
et al. (2007a) and empty red circles are stars with GIRAFFE spectra studied in
Carretta et al. (2009b).}
\label{f:cmdu63al}
\end{figure}

For sake of homogeneity with our previous work, we used our $BVI$ photometry
obtained with the Wide Field Imager (WFI) at the 2.2 m ESO/MPI telescope in La
Silla  (ESO Programme 69.D-0582). The detailed reduction and calibration
procedures  for photometry and astrometry are provided in Carretta et al.
(2007a). Optical photometry was complemented with near IR $K$ magnitudes from
2MASS (Skrutskie et al. 2006), mandatory to obtain the atmospheric parameters
onto our homogeneous scale (see below).  Stars A10 and A14 have no
identifications in our WFI photometry; $V$ magnitudes from these objects were
assigned using a $V$ vs $K$ calibration derived from  other 47 stars in NGC~6388
(Carretta et al. 2007a, 2009b) with both sets of data available. The new sample,
as well as the previously analysed datasets, are shown on the $V,B-V$
colour-magnitude diagram (CMD) in  Fig.~\ref{f:cmdu63al}.

The UVES Red arm spectra (obtained with the 580nm setup, spectral range $\sim 4800-6800$~\AA, 
resolution 
$\sim 47,000$) were reduced with the ESO pipeline using Reflex v2.8.5 . 
The de-biased, flat-fielded, extracted to 1-d and wavelength calibrated spectra
were sky  subtracted and  shifted to zero radial velocity using
IRAF\footnote{IRAF is the Image Reduction and Analysis Facility, a general
purpose software system for the reduction and analysis of astronomical data.
IRAF is written and supported by the IRAF programming group at the National
Optical Astronomy Observatories (NOAO) in Tucson, Arizona. NOAO is operated by
the Association of Universities for Research in Astronomy (AURA), Inc. under
cooperative agreement with the National Science Foundation.}. 
We measured RVs for each 
spectrum using about 80 atomic lines with the IRAF package
{\sc rvidlines}. 
The resulting heliocentric RVs and the relative errors are shown in the last columns of
Table~\ref{t:sample}.
From our sample of 17 stars we found an average RV of $77.7\pm 2.9$,
$\sigma=11.8$ kms$^{-1}$, in good agreement with the values derived in Carretta
(2007a) from seven stars only ($79.1\pm 1.0$ $\sigma=3.0$ kms$^{-1}$ and in the
literature ($81.2\pm 1.2$ kms$^{-1}$, Harris 1996, and $82.0\pm 0.5$ $\sigma=7.7$ 
kms$^{-1}$ from 240 stars in L13).
The spectra were then shifted to zero radial velocity and coadded
for each star.

\section{Analysis, atmospheric parameters and metallicity}

The steps for the derivation of the atmospheric parameters were the same adopted
in all our FLAMES survey, and in particular for the analysis of the seven giants with
UVES spectra in NGC~6388 (Carretta et al. 2007a). Effective temperatures are
obtained from a calibration between $K$ magnitudes and $T_{\rm eff}$ from dereddened 
$V-K$ (using the Alonso et al. 1999, 2001 relation). The calibration as a
function of $K$ magnitudes was derived from 33 member stars in NGC~6388
(Carretta et al. 2009b). Derived temperatures were used with apparent
magnitudes, a distance modulus $(m-M)_V=16.14$, and bolometric corrections from
Alonso et al. (1999) to derive surface gravities (adopting masses of 0.90
$M_\odot$ and $M_{bol,\odot}=4.75$. Distance modulus, and the adopted reddening 
$E(B - V) = 0.37$ are from Harris (1996). We used the relations 
$E(V - K) = 2.75E(B - V),  A_V = 3.1E(B - V)$, and
$A_K = 0.353E(B - V)$ from Cardelli et al. (1989).	

Our analysis is based on equivalent width ($EW$s) measured with the ROSA package
(Gratton 1988) following the procedure described in Bragaglia et al. (2001).
Line list and solar reference abundances, used throughout our FLAMES survey, 
are described in Gratton et al. (2003). Values of the microturbulent velocity
$v_t$ were derived by minimizing the slope of the relation between abundances of
Fe~{\sc i} and expected line strength (see Magain 1984). Finally, the model
atmosphere with appropriated parameters is chosen from the Kurucz (1993) grid by
selecting the model with abundance equal to the average abundance from 
Fe~{\sc i} lines.

\begin{table*}
\centering
\caption[]{Adopted atmospheric parameters and derived iron abundances in
NGC~6388}
\begin{tabular}{rccccrccrcc}
\hline
Star   &  $T_{\rm eff}$ & $\log$ $g$ & [A/H]  &$v_t$	     & nr & [Fe/H]{\sc i} & $rms$ & nr & [Fe/H{\sc ii} & $rms$ \\
       &     (K)	&  (dex)     & (dex)  &(km s$^{-1}$) &    & (dex)	  &	  &    & (dex)         &       \\
\hline
A01  & 3773 & 0.78 & $-$0.47 & 1.87 &  79 & $-$0.467 & 0.164 &  5  & $-$0.414 & 0.121  \\
A02  & 3738 & 0.68 & $-$0.54 & 1.51 &  73 & $-$0.537 & 0.161 & 15  & $-$0.433 &	0.219  \\
A03  & 3979 & 1.31 & $-$0.45 & 2.02 & 124 & $-$0.454 & 0.146 & 17  & $-$0.284 &	0.173  \\
A04  & 3727 & 0.65 & $-$0.47 & 1.70 &  95 & $-$0.470 & 0.160 & 14  & $-$0.349 & 0.165  \\
A05  & 3836 & 0.97 & $-$0.52 & 1.57 & 108 & $-$0.520 & 0.126 & 17  & $-$0.401 & 0.205  \\
A06  & 3793 & 0.85 & $-$0.53 & 1.54 &  89 & $-$0.531 & 0.120 & 17  & $-$0.454 & 0.212  \\
A07  & 3758 & 0.75 & $-$0.53 & 1.52 & 101 & $-$0.529 & 0.204 & 11  & $-$0.488 & 0.175  \\
A08  & 3779 & 0.81 & $-$0.47 & 1.53 & 102 & $-$0.461 & 0.177 &  8  & $-$0.353 &	0.149  \\
A09  & 3925 & 1.20 & $-$0.44 & 1.63 & 104 & $-$0.439 & 0.144 & 15  & $-$0.264 &	0.173  \\
A10  & 3834 & 0.92 & $-$0.50 & 1.09 & 100 & $-$0.499 & 0.223 & 10  & $-$0.433 &	0.237  \\
A11  & 3874 & 1.07 & $-$0.44 & 1.70 & 109 & $-$0.439 & 0.166 &  9  & $-$0.336 &	0.210  \\
A12  & 4256 & 1.76 & $-$0.46 & 1.86 & 125 & $-$0.464 & 0.164 & 18  & $-$0.359 &	0.145  \\
A13  & 4065 & 1.43 & $-$0.43 & 1.65 & 120 & $-$0.430 & 0.149 & 13  & $-$0.369 &	0.113  \\
A14  & 4247 & 1.76 & $-$0.48 & 1.34 & 127 & $-$0.478 & 0.156 & 25  & $-$0.324 &	0.188  \\
A15  & 4332 & 1.89 & $-$0.40 & 1.80 &  89 & $-$0.401 & 0.134 &  8  & $-$0.403 &	0.198  \\
A16  & 4197 & 1.67 & $-$0.48 & 1.87 & 100 & $-$0.479 & 0.150 &  8  & $-$0.385 &	0.138  \\
A17  & 4223 & 1.71 & $-$0.47 & 1.78 & 116 & $-$0.473 & 0.192 & 16  & $-$0.374 &	0.252  \\
\hline
\end{tabular}
\label{t:atmpar63al}
\end{table*}

Derived atmospheric parameters and Fe abundances for individual stars are listed in
Table~\ref{t:atmpar63al} whereas mean abundances for the new sample, the previous
analysis by Carretta et al. (2007a) and the combined sample are shown in
Table~\ref{t:meanabu63al}. 

\begin{table*}
\centering
\caption{Mean abundances from UVES for the present and previous samples and their combination }
\begin{tabular}{lrcccrcccrccc}
\hline
 Element             &stars &  avg.&  rms & ref.&stars &  avg.&  rms & ref.&stars &  avg.&  rms & ref. \\
\hline

$[$Fe/H$]${\sc i}    & 17 & $-$0.475 & 0.038 & (1) & 7 & $-$0.441 & 0.038 & (2) & 24 & $-$0.465 & 0.041 & (1) \\ 
$[$Fe/H$]${\sc ii}   & 17 & $-$0.378 & 0.059 & (1) & 7 & $-$0.368 & 0.088 & (2) & 24 & $-$0.375 & 0.067 & (1) \\ 
$[$O/Fe$]${\sc i}    & 17 & $-$0.167 & 0.201 & (1) & 7 & $-$0.299 & 0.159 & (2) & 24 & $-$0.206 & 0.196 & (1) \\ 
$[$Na/Fe$]${\sc i}   & 17 &   +0.469 & 0.150 & (1) & 7 &   +0.595 & 0.156 & (2) & 24 &   +0.506 & 0.160 & (1) \\ 
$[$Mg/Fe$]${\sc i}   & 17 &   +0.216 & 0.048 & (1) & 7 &   +0.208 & 0.066 & (2) & 24 &   +0.214 & 0.053 & (1) \\ 
$[$Al/Fe$]${\sc i}   & 17 &   +0.347 & 0.308 & (1) & 7 &   +0.688 & 0.243 & (2) & 24 &   +0.447 & 0.327 & (1) \\
$[$Si/Fe$]${\sc i}   & 17 &   +0.369 & 0.056 & (1) & 7 &   +0.322 & 0.102 & (2) & 24 &   +0.356 & 0.073 & (1) \\ 
$[$Ca/Fe$]$ {\sc i}  & 17 &   +0.046 & 0.039 & (1) & 7 &   +0.064 & 0.062 & (2) & 24 &   +0.051 & 0.046 & (1) \\ 
$[$Ti/Fe$]${\sc i}   & 17 &   +0.265 & 0.083 & (1) & 7 &   +0.367 & 0.099 & (2) & 24 &   +0.294 & 0.098 & (1) \\ 
$[$Ti/Fe$]${\sc ii}  & 17 &   +0.187 & 0.068 & (1) & 7 &   +0.299 & 0.117 & (2) & 24 &   +0.220 & 0.097 & (1) \\ 
$[$Sc/Fe$]${\sc ii}  & 17 & $-$0.066 & 0.081 & (1) & 7 &   +0.050 & 0.065 & (2) & 24 & $-$0.032 & 0.093 & (1) \\ 
$[$V/Fe$]${\sc i}    & 17 &   +0.233 & 0.155 & (1) & 7 &   +0.390 & 0.099 & (2) & 24 &   +0.278 & 0.157 & (1) \\ 
$[$Cr/Fe$]${\sc i}   & 17 & $-$0.105 & 0.074 & (1) & 7 & $-$0.037 & 0.109 & (2) & 24 & $-$0.085 & 0.089 & (1) \\ 
$[$Mn/Fe$]${\sc i}   & 17 & $-$0.204 & 0.048 & (1) & 7 & $-$0.248 & 0.024 & (2) & 24 & $-$0.217 & 0.047 & (1) \\ 
$[$Co/Fe$]${\sc i}   & 17 &   +0.038 & 0.088 & (1) & 7 &   +0.042 & 0.075 & (2) & 24 &   +0.039 & 0.083 & (1) \\ 
$[$Ni/Fe$]${\sc i}   & 17 &   +0.034 & 0.033 & (1) & 7 &   +0.034 & 0.033 & (2) & 24 &   +0.034 & 0.032 & (1) \\ 
$[$Zn/Fe$]${\sc i}   & 16 &   +0.073 & 0.282 & (1) & 7 &   +0.128 & 0.254 & (1) & 23 &   +0.090 & 0.269 & (1) \\ 
\hline
\end{tabular}
\begin{list}{}{}
\item[(1)] this work
\item[(2)] Carretta et al. (2007a)
\end{list}
\label{t:meanabu63al}
\end{table*}

\section{Errors analysis and derived metallicity}
The procedure to estimate star to star errors due to uncertainties in the 
adopted atmospheric parameters and in the $EW$ measurements closely follows step
by step the one in Carretta et al. (2007a), so we will not further describe it here.

\begin{table*}
\centering
\caption[]{Sensitivities of abundance ratios to variations in the atmospheric
parameters and to errors in the equivalent widths, and errors in abundances for
stars of NGC~6388 observed with UVES.}
\begin{tabular}{lrrrrrrrr}
\hline
Element     & Average   & T$_{\rm eff}$ & $\log g$ & [A/H]   & $v_t$    & EWs     & Total   & Total      \\
            & n. lines  &      (K)      &  (dex)   & (dex)   &kms$^{-1}$& (dex)   &Internal & Systematic \\
\hline        
Variation&              &  50           &   0.20   &  0.10   &  0.10    &         &         &            \\
Internal &              &   6           &   0.04   &  0.04   &  0.09    & 0.161   &         &            \\
Systematic&             &  35           &   0.06   &  0.02   &  0.02    &         &         &            \\
\hline
$[$Fe/H$]${\sc  i}& 104 &  $-$0.006	&   +0.040 &  +0.024 & $-$0.045 & 0.016   &0.045    &0.018	 \\
$[$Fe/H$]${\sc ii}&  13 &  $-$0.097	&   +0.122 &  +0.043 & $-$0.033 & 0.045   &0.063    &0.078	 \\
$[$O/Fe$]${\sc  i}&   2 &    +0.023	&   +0.041 &  +0.015 &   +0.044 & 0.114   &0.121    &0.054	 \\
$[$Na/Fe$]${\sc i}&   4 &    +0.047	& $-$0.086 &  +0.010 &   +0.002 & 0.081   &0.083    &0.055	 \\
$[$Mg/Fe$]${\sc i}&   4 &  $-$0.002	& $-$0.034 &$-$0.002 &   +0.016 & 0.081   &0.082    &0.016	 \\
$[$Al/Fe$]${\sc i}&   2 &    +0.044	& $-$0.039 &$-$0.022 &   +0.013 & 0.114   &0.115    &0.182	 \\
$[$Si/Fe$]${\sc i}&   8 &  $-$0.044	&   +0.019 &$-$0.001 &   +0.026 & 0.057   &0.062    &0.035	 \\
$[$Ca/Fe$]${\sc i}&  18 &    +0.067	& $-$0.070 &$-$0.016 & $-$0.023 & 0.038   &0.047    &0.052	 \\
$[$Sc/Fe$]${\sc ii}&  8 &    +0.085	& $-$0.036 &$-$0.007 & $-$0.016 & 0.057   &0.060    &0.064	 \\
$[$Ti/Fe$]${\sc i}&  22 &    +0.090	& $-$0.045 &$-$0.012 & $-$0.039 & 0.034   &0.051    &0.068	 \\
$[$Ti/Fe$]${\sc ii}& 10 &    +0.072	& $-$0.038 &$-$0.010 & $-$0.024 & 0.051   &0.057    &0.054	 \\
$[$V/Fe$]${\sc i} &  11 &    +0.096	& $-$0.038 &$-$0.008 & $-$0.030 & 0.049   &0.057    &0.078	 \\
$[$Cr/Fe$]${\sc i}&  24 &    +0.053	& $-$0.042 &$-$0.015 & $-$0.004 & 0.033   &0.035    &0.043	 \\
$[$Mn/Fe$]${\sc i}&   7 &    +0.041	& $-$0.032 &  +0.004 & $-$0.013 & 0.061   &0.063    &0.033	 \\
$[$Co/Fe$]${\sc i}&   5 &  $-$0.006	&   +0.008 &$-$0.003 & $-$0.004 & 0.072   &0.072    &0.022	 \\
$[$Ni/Fe$]${\sc i}&  38 &  $-$0.010	&   +0.016 &$-$0.002 &   +0.008 & 0.026   &0.027    &0.012	 \\
$[$Zn/Fe$]${\sc i}&   1 &  $-$0.033	&   +0.011 &$-$0.000 & $-$0.011 & 0.161   &0.161    &0.074	 \\

\hline
\end{tabular}
\label{t:sensitivityu63al}
\end{table*}

Results are summarized in Table~\ref{t:sensitivityu63al} where
we show the adopted variation and
estimated errors in the atmospheric parameters (first three lines),
the sensitivities of abundances to changes in atmospheric parameters (body of the
Table, columns 3 to 4) and to errors in $EW$s (column 6), and the estimated
star-to-star and systematic errors (last 2 columns).
The main difference with the treatment of errors as described in Carretta et al.
(2007a) is that sensitivities are derived not from a single star in the middle
of the temperature range, but instead using averages from all stars in the
sample. 
Other differences in the adopted uncertainties are those related to 
quantities that vary in the two analysis: number of lines (affecting the
internal error in $EW$ measurements), the slope of the relation
abundance-expected line strength (affecting the internal error in $v_t$), the
mean dereddened $V-K$ colour (which enters in the systematic error in 
$T_{\rm eff}$) and so on.

On the abundance scale defined in our FLAMES survey, the metallicity of NGC~6388
from the present sample with UVES spectra is [Fe/H]$=-0.475\pm0.009\pm0.018$ dex, $\sigma=0.038$
dex (17 stars), where the first and the second are statistical and systematic
errors, respectively. We do not find evidence of intrinsic scatter in the
metal-abundance for NGC~6388, since the observed scatter $0.038\pm0.009$ dex
well agrees with the scatter expected by the uncertainties in the analysis,
$0.045\pm0.011$ (see Table~\ref{t:sensitivityu63al}). This conclusion is
strengthened by the combined sample with UVES spectra: when we add the seven
stars analysed in a homogeneous way in Carretta et al. (2007a), we obtain a mean
metallicity [Fe/H]$=-0.465\pm0.008$ dex, with $\sigma=0.041$ dex (24 stars),
excluding any intrinsic dispersion in Fe in this cluster.

\begin{figure}
\centering
\includegraphics[scale=0.40]{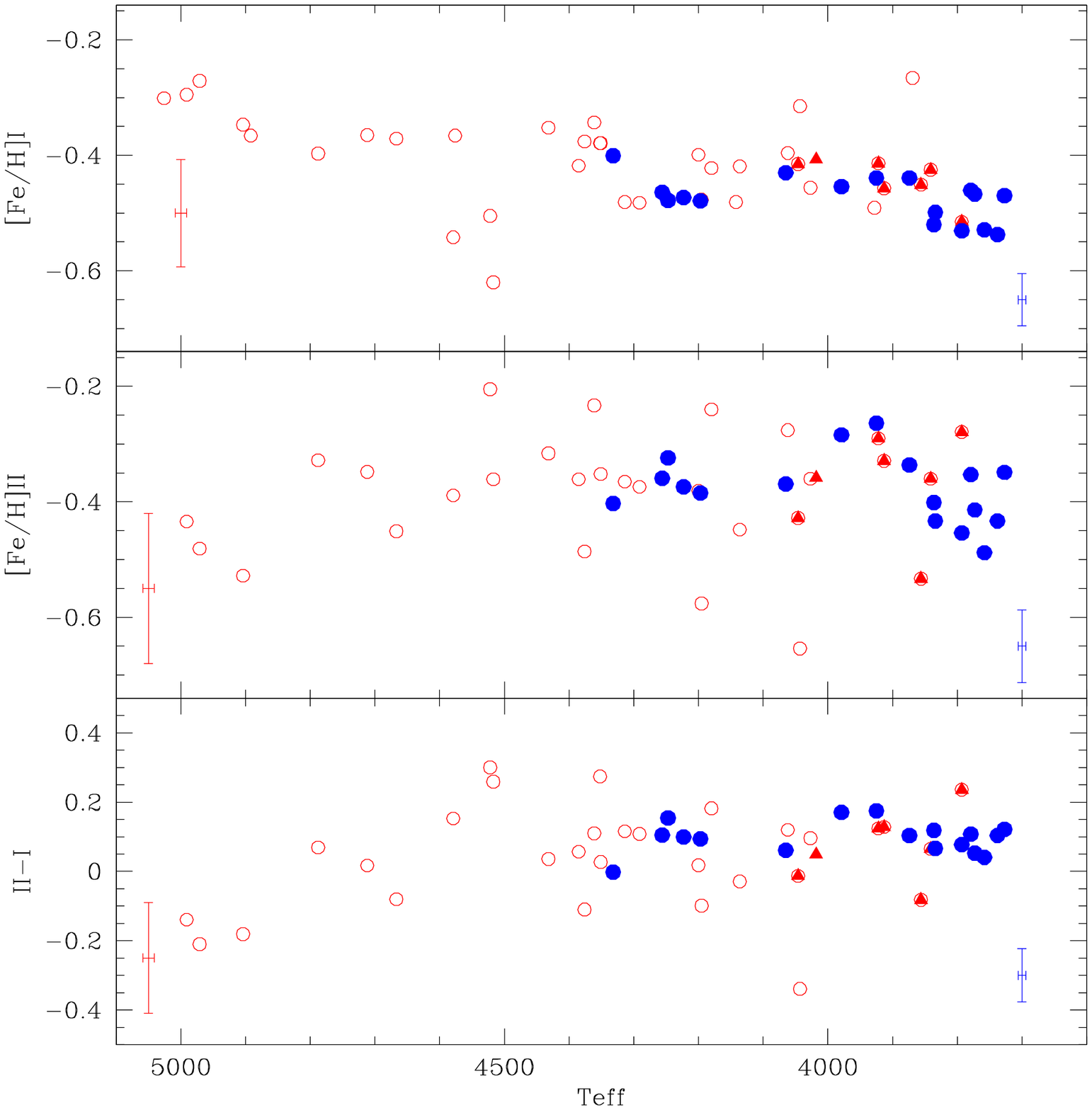}
\caption{Fe abundances from neutral and singly ionized lines (upper and middle
panels, respectively) and their differences (lower panel) as a function of the
effective temperatures. Filled symbols are stars with UVES spectra (blue
circles: this work, red triangles: Carretta et al. 2007a), whereas open circles
are stars with GIRAFFE spectra (Carretta et al. 2009b). Star-to-star error bars
for GIRAFFE and UVES are indicated on the left and on the right side,
respectively.}
\label{f:feteff3}
\end{figure}

Abundances of Fe from neutral and singly ionized lines are shown as a function
of the effective temperatures in Fig.~\ref{f:feteff3}, upper and lower panels,
respectively, where we compare results from UVES spectra (current and previous
work, filled squares and triangles) to those obtained from GIRAFFE spectra 
(Carretta et al. 2009b, open circles). The 
difference [Fe/H]~{\sc ii} - [Fe/H]~{\sc i} is plotted in the lower panel.
As in Carretta et al. (2007a), the mean value from singly ionized Fe lines is
slighly higher: [Fe/H]$=-0.378\pm0.014$ dex, $\sigma=0.059$ dex (17 stars).
On average, the difference is scarcely significant: $0.097\pm0.011$ dex,
$\sigma= 0.046$ dex for the present sample 
($0.090\pm0.014$ dex, $\sigma= 0.066$ dex for the combined sample of 24 stars
with UVES spectra). This offset is most likely related to uncertainties in the
measurement of relatively weak lines on spectra of moderate S/N. We can exclude
deviations from the local thermodynamic equilibrium (LTE) since we do not
observe any trend as a function of temperature. Should the difference be due to
departures from LTE, we would expect larger effects in 
lower gravity (lower $T_{\rm eff}$) stars, where the thermalizing effect of
collisions is not pronounced (Gratton et al. 1999), which is not the case (lower
panel in Fig.~\ref{f:feteff3}).

Abundances for proton-capture, $\alpha-$capture, and iron-peak 
elements in individual RGB stars of NGC~6388 from the present analysis are
listed in Tab.~\ref{t:protonu63al}, Tab.~\ref{t:alphau63al}, and
Tab.~\ref{t:fepeaku63al}, respectively.
Average values are in Tab.~\ref{t:meanabu63al}. Finally, in Tab.~\ref{t:zn}
we report abundances of Zn obtained from UVES spectra of the 7 stars
analyzed in Carretta et al. (2007a), since this species was not derived in
that work.

\begin{table*}
\centering
\caption[]{Light element abundances.}
\begin{tabular}{rccccrcccrcrc}
\hline
Star   &   nr &  [O/Fe]{\sc i} &  rms &  nr &  [Na/Fe]{\sc i} &  rms &  nr &  [Mg/Fe]{\sc i} &  rms  &  nr &  [Al/Fe]{\sc i} &  rms  \\
\hline
A01    &   1  & $-$0.539       &      &  3  &   +0.620        & 0.165&  4  &       +0.280    &  0.056&  2  &    +1.007       &  0.177 \\
A02    &   2  & $-$0.216       & 0.115&  4  &	+0.675	      & 0.093&	4  &       +0.267    &  0.180&	2  &    +0.416       &  0.004 \\
A03    &   1  & $-$0.444       &      &  4  &	+0.478	      & 0.054&	4  &       +0.171    &  0.177&	2  &    +0.502       &  0.151 \\
A04    &   2  & $-$0.419       & 0.029&  4  &	+0.665	      & 0.078&	3  &       +0.176    &  0.158&	2  &    +0.532       &  0.173 \\
A05    &   2  & $-$0.343       & 0.139&  4  &	+0.676	      & 0.080&	4  &       +0.201    &  0.063&	2  &    +0.591       &  0.168 \\
A06    &   2  & $-$0.064       & 0.059&  4  &	+0.571	      & 0.041&	4  &       +0.223    &  0.217&	2  &    +0.211       &  0.035 \\
A07    &   2  & $-$0.091       & 0.011&  4  &	+0.504	      & 0.072&	4  &       +0.241    &  0.134&	2  &    +0.256       &  0.110 \\
A08    &   2  & $-$0.230       & 0.069&  4  &	+0.555	      & 0.087&	4  &       +0.273    &  0.039&	2  &    +0.764       &  0.265 \\
A09    &   2  & $-$0.096       & 0.079&  4  &	+0.436	      & 0.093&	4  &       +0.198    &  0.239&	2  &  $-$0.052       &  0.086 \\
A10    &   2  & $-$0.011       & 0.049&  4  &	+0.331	      & 0.110&	3  &       +0.301    &  0.260&	2  &    +0.053       &  0.000 \\
A11    &   2  &   +0.078       & 0.021&  4  &	+0.282	      & 0.154&	4  &       +0.237    &  0.141&	2  &  $-$0.035       &  0.106 \\
A12    &   1  & $-$0.019       &      &  4  &	+0.389	      & 0.103&	3  &       +0.157    &  0.103&	2  &    +0.491       &  0.006 \\
A13    &   1  & $-$0.126       &      &  3  &	+0.316	      & 0.133&	4  &       +0.160    &  0.147&	2  &    +0.034       &  0.131 \\
A14    &   1  & $-$0.072       &      &  4  &	+0.304	      & 0.113&	4  &       +0.210    &  0.018&	2  &    +0.014       &  0.327 \\
A15    &   1  & $-$0.167       &      &  3  &	+0.483	      & 0.054&	2  &       +0.240    &  0.185&	2  &    +0.621       &  0.093 \\
A16    &   1  &   +0.242       &      &  4  &	+0.182	      & 0.172&	3  &       +0.214    &  0.167&	1  &    +0.096       &  9.999 \\
A17    &   1  & $-$0.326       &      &  3  &	+0.502	      & 0.119&	4  &       +0.129    &  0.143&	2  &    +0.404       &  0.261 \\

\hline
\end{tabular}
\label{t:protonu63al}
\end{table*}

\begin{table*}
\centering
\caption[]{Abundances of $\alpha$ elements.}
\begin{tabular}{rccccrcccrccc}
\hline
Star   &   nr &  [Si/Fe]{\sc i} &  rms  & nr &  [Ca/Fe]{\sc i} &  rms  &  nr &  [Ti/Fe]{\sc i} &  rms   &  nr&  [Ti/Fe]{\sc ii} &  rms  \\
\hline
A01    &   6  &    +0.358	& 0.041 & 15 &       +0.066    & 0.140 &  21 &   +0.364        &  0.248 &  6 &      +0.237      & 0.219 \\
A02    &   8  &    +0.261	& 0.183 & 19 &       +0.054    & 0.194 &  14 &   +0.385        &  0.212 &  8 &      +0.104      & 0.163 \\
A03    &   8  &    +0.436	& 0.119 & 15 &     $-$0.014    & 0.189 &  19 &   +0.083        &  0.119 & 10 &      +0.005      & 0.149 \\
A04    &   6  &    +0.397	& 0.198 & 18 &       +0.030    & 0.215 &  15 &   +0.259        &  0.224 &  7 &      +0.130      & 0.155 \\
A05    &   8  &    +0.381	& 0.203 & 15 &       +0.093    & 0.200 &  20 &   +0.376        &  0.191 &  8 &      +0.158      & 0.156 \\
A06    &   7  &    +0.283	& 0.084 & 17 &       +0.043    & 0.175 &  26 &   +0.291        &  0.264 & 11 &      +0.225      & 0.196 \\
A07    &   8  &    +0.328	& 0.206 & 18 &       +0.083    & 0.175 &  14 &   +0.375        &  0.231 & 11 &      +0.271      & 0.236 \\
A08    &   8  &    +0.353	& 0.267 & 15 &       +0.090    & 0.201 &  14 &   +0.234        &  0.183 & 12 &      +0.223      & 0.304 \\
A09    &   9  &    +0.423	& 0.165 & 17 &       +0.078    & 0.176 &  20 &   +0.221        &  0.202 &  6 &      +0.127      & 0.209 \\
A10    &   8  &    +0.369	& 0.236 & 21 &     $-$0.024    & 0.211 &  17 &   +0.217        &  0.304 &  9 &      +0.219      & 0.250 \\
A11    &   6  &    +0.381	& 0.103 & 19 &       +0.012    & 0.190 &  25 &   +0.239        &  0.297 &  9 &      +0.174      & 0.216 \\
A12    &   8  &    +0.388	& 0.142 & 20 &       +0.089    & 0.188 &  32 &   +0.243        &  0.171 & 11 &      +0.252      & 0.200 \\
A13    &   8  &    +0.388	& 0.124 & 20 &       +0.023    & 0.210 &  21 &   +0.125        &  0.181 & 12 &      +0.171      & 0.206 \\
A14    &   8  &    +0.270	& 0.142 & 19 &       +0.064    & 0.204 &  27 &   +0.258        &  0.206 & 11 &      +0.183      & 0.168 \\
A15    &   7  &    +0.376	& 0.187 & 17 &       +0.042    & 0.168 &  28 &   +0.269        &  0.172 & 10 &      +0.250      & 0.264 \\
A16    &   8  &    +0.447	& 0.249 & 18 &     $-$0.022    & 0.179 &  19 &   +0.246        &  0.247 & 11 &      +0.205      & 0.257 \\
A17    &   8  &    +0.438	& 0.267 & 20 &       +0.072    & 0.222 &  34 &   +0.315        &  0.319 & 11 &      +0.249      & 0.224 \\

\hline
\end{tabular}
\label{t:alphau63al}
\end{table*}

\begin{table*}
\centering
\setlength{\tabcolsep}{1mm}
\caption[]{Iron-peak abundances.}
\small
\begin{tabular}{rccccrcccrcccrcccrccc}
\hline
Star   
&   nr &  [Sc/Fe]{\sc ii} &  rms &  nr &  [V/Fe]{\sc i} &  rms &  nr &  [Cr/Fe]{\sc i} &  rms &     nr &  [Mn/Fe]{\sc  i} &  rms &  nr &  [Co/Fe]{\sc i}&  rms &  nr &  [Ni/Fe]{\sc i} &  rms &    nr &  [Zn/Fe]{\sc i} \\
\hline
A01    & 5  &	 +0.098  &  0.123   &   13  &	0.325	& 0.186  &   22  & $-$0.067 &	0.215	&  6  & $-$0.244  &   0.243  &    4 &  +0.077 &   0.188 &   32  &   +0.044 &   0.155 &  1  &$-$0.224   	 \\
A02    & 8  &  $-$0.122  &  0.162   &   10  &	0.419	& 0.190  &   21  &   +0.001 &	0.258	&  6  & $-$0.109  &   0.148  &	  5 &  +0.055 &   0.129 &   38  &   +0.016 &   0.237 &  1  &  +0.194   	 \\
A03    & 8  &  $-$0.229  &  0.101   &   13  &$-$0.072	& 0.296  &   24  & $-$0.269 &	0.184	&  8  & $-$0.236  &   0.181  &	  5 &$-$0.104 &   0.159 &   45  &   +0.012 &   0.185 &  1  &$-$0.036   	 \\
A04    & 8  &  $-$0.016  &  0.142   &	 9  &	0.366	& 0.122  &   17  & $-$0.085 &	0.214	&  6  & $-$0.213  &   0.266  &	  5 &  +0.011 &   0.099 &   35  & $-$0.013 &   0.186 &  1  &  +0.047  	 \\
A05    & 8  &	 +0.010  &  0.089   &	 9  &	0.416	& 0.122  &   26  & $-$0.025 &	0.242	&  6  & $-$0.149  &   0.138  &	  5 &  +0.132 &   0.154 &   41  &   +0.070 &   0.175 &  1  &  +0.119  	 \\
A06    & 8  &  $-$0.046  &  0.102   &   10  &	0.367	& 0.146  &   22  & $-$0.083 &	0.223	&  5  & $-$0.216  &   0.172  &	  5 &  +0.038 &   0.130 &   38  & $-$0.013 &   0.147 &  1  &  +0.111   	 \\
A07    & 8  &  $-$0.069  &  0.109   &   13  &	0.361	& 0.137  &   26  & $-$0.069 &	0.294	&  7  & $-$0.129  &   0.261  &	  5 &  +0.071 &   0.181 &   38  &   +0.014 &   0.201 &  1  &  +0.234   	 \\
A08    & 8  &  $-$0.078  &  0.269   &   13  &	0.219	& 0.134  &   18  & $-$0.077 &	0.204	&  6  & $-$0.259  &   0.086  &	  5 &  +0.045 &   0.094 &   34  &   +0.048 &   0.176 &  1  &  +0.076   	 \\
A09    & 8  &  $-$0.066  &  0.138   &	 9  &	0.154	& 0.244  &   21  & $-$0.198 &	0.168	&  6  & $-$0.239  &   0.212  &	  5 &  +0.045 &   0.110 &   39  &   +0.018 &   0.135 &  1  &  +0.434   	 \\
A10    & 8  &  $-$0.128  &  0.180   &   12  &	0.199	& 0.166  &   25  & $-$0.050 &	0.385	&  6  & $-$0.219  &   0.212  &	  5 &  +0.196 &   0.270 &   36  &   +0.039 &   0.255 &  1  &$-$0.341   	 \\
A11    & 8  &  $-$0.122  &  0.173   &   12  &	0.146	& 0.179  &   23  & $-$0.162 &	0.238	&  7  & $-$0.159  &   0.194  &	  5 &$-$0.023 &   0.264 &   37  &   +0.063 &   0.215 &  1  &  +0.096   	 \\
A12    & 8  &	 +0.004  &  0.223   &   13  &	0.130	& 0.337  &   27  & $-$0.087 &	0.215	&  8  & $-$0.239  &   0.260  &	  4 &  +0.021 &   0.200 &   41  &   +0.032 &   0.223 &  1  &  +0.551   	 \\
A13    & 8  &  $-$0.113  &  0.138   &   11  &$-$0.012	& 0.256  &   25  & $-$0.190 &	0.194	&  6  & $-$0.163  &   0.068  &	  3 &$-$0.061 &   0.185 &   43  &   +0.014 &   0.137 &  1  &  +0.235   	 \\
A14    & 8  &  $-$0.153  &  0.166   &   12  &	0.238	& 0.317  &   27  & $-$0.124 &	0.212	&  8  & $-$0.166  &   0.186  &	  5 &$-$0.076 &   0.129 &   44  & $-$0.001 &   0.186 &  1  &$-$0.205   	 \\
A15    & 8  &  $-$0.013  &  0.247   &   12  &	0.292	& 0.340  &   27  & $-$0.029 &	0.208	&  7  & $-$0.243  &   0.245  &	  5 &  +0.170 &   0.249 &   31  &   +0.065 &   0.240 &  1  &  +0.373   	 \\
A16    & 8  &  $-$0.124  &  0.245   &   10  &	0.013	& 0.323  &   22  & $-$0.076 &	0.302	&  7  & $-$0.221  &   0.100  &	  3 &  +0.134 &   0.270 &   37  &   +0.103 &   0.286 &     &	      	 \\
A17    & 8  &	 +0.045  &  0.203   &   11  &	0.392	& 0.279  &   27  & $-$0.200 &	0.233	&  7  & $-$0.265  &   0.225  &	  5 &$-$0.078 &   0.086 &   38  &   +0.069 &   0.198 &  1  &$-$0.493   	 \\

\hline
\end{tabular}
\label{t:fepeaku63al}
\end{table*}

\section{Multiple stellar populations}

\subsection{Proton-capture elements in NGC~6388}

We derived the abundances of O, Na, Mg, and Al, light elements involved in the
network of proton-capture reactions in hot H-burning, for all 17 stars in the
present sample. 

The forbidden line [O~{\sc i}] 6300~\AA\ was measured on the spectra after cleaning 
from contamination by telluric lines. Using the coolest star in our sample, we 
checked that the blending of the weak, high excitation Ni~{\sc i} line at 
6300.336~\AA\ is expected  to contribute about 4.4 m\AA\ to the $EW$ of 
the [O~{\sc i}] line, using the
line parameters by Johansson et al. (2003). This corresponds to
correcting the O abundances about 0.031 dex downward. On the other hand, this
effect is almost compensated by the underestimate of O abundances due to
neglecting CO formation, so that we followed the approach used by Gratton et 
al. (2006) in the analysis of NGC~6441, a twin GC of NGC~6388 in several
features, and we did not apply any corrections for the above (small) effects.

Na abundances are corrected for effects of departures from the LTE assumption
using prescriptions by Gratton et al. (1999) as done homogeneously in our group's
FLAMES survey.

\begin{figure}
\centering
\includegraphics[scale=0.40]{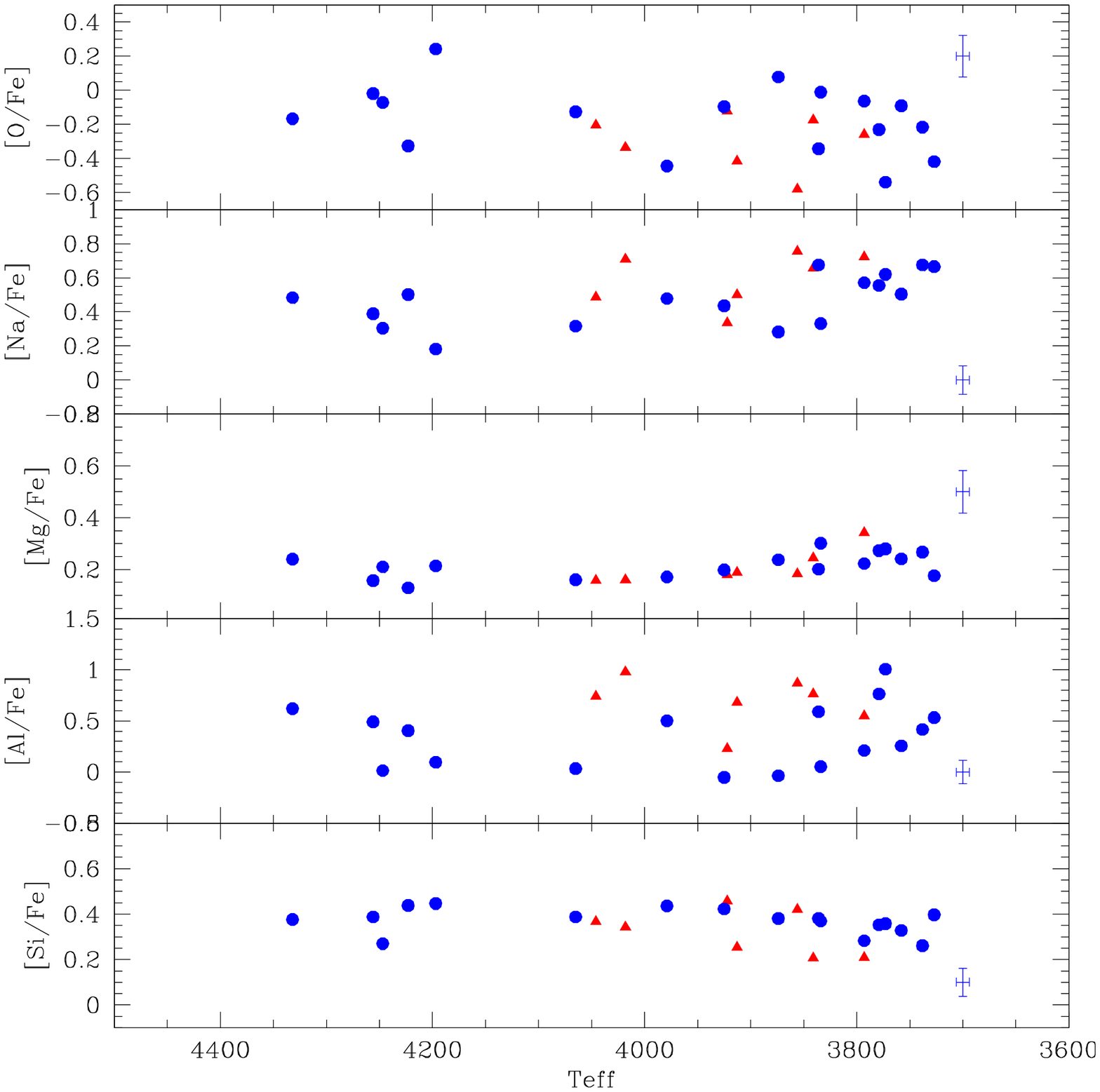}
\caption{Abundance ratios for proton-capture elements O, Na, Mg, Al, and Si
(from top to bottom) from UVES spectra as a function of the effective
temperatures in NGC~6388. Red triangles: RGB stars from Carretta et al. (2007a).
Blue circles: new sample from present work, with internal errors plotted in each
panel.}
\label{f:protonteff}
\end{figure}

\begin{figure}
\centering
\includegraphics[scale=0.42]{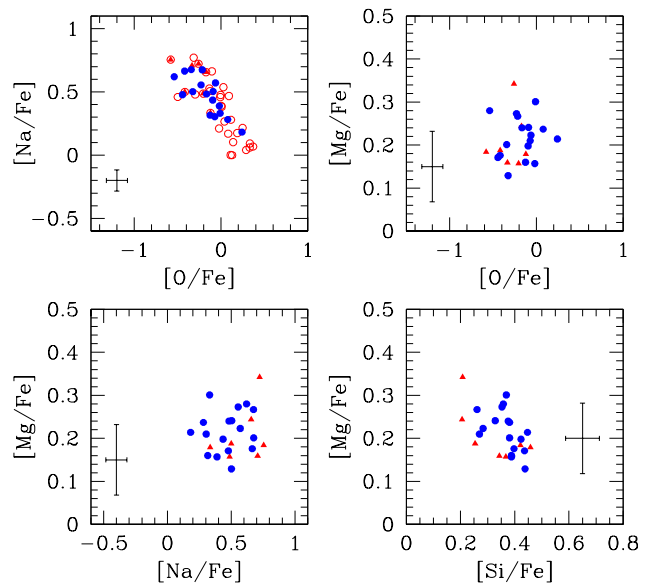}
\caption{Relations among the abundance ratios of O, Na, Mg, and Si. Symbols are
as in Fig.~\ref{f:protonteff}. In the top-left panel, open circles are stars of
NGC~6388 with abundances derived from GIRAFFE spectra in Carretta et al.
(2009b).}
\label{f:checku63al1}
\end{figure}

\begin{figure}
\centering
\includegraphics[scale=0.42]{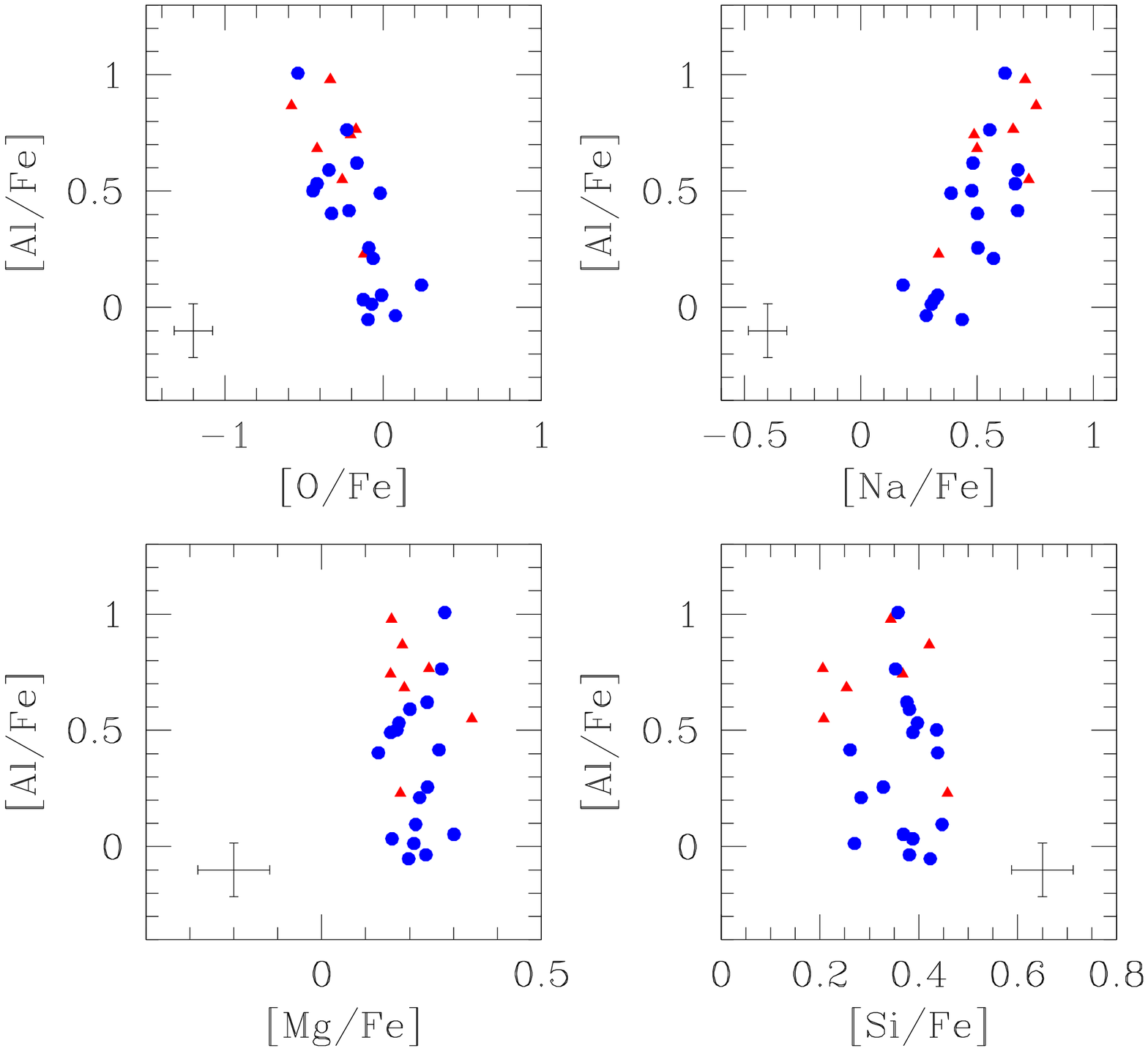}
\caption{[Al/Fe] as a function of O, Na, Mg, and Si. Symbols are as in
Fig.~\ref{f:protonteff}}
\label{f:checku63al2}
\end{figure}

The run of abundance ratios for O, Na, Mg, Al, and Si as a function of effective
temperatures is shown in Fig.~\ref{f:protonteff}, whereas in 
Fig.~\ref{f:checku63al1} and Fig.~\ref{f:checku63al2} we summarize the 
relations among these light elements. In the present work we more than tripled 
(from 7 to 24) the number of stars with proton-capture elements derived from
high-resolution UVES spectra. Moreover, in the Na-O plane we could add also
stars with abundances obtained with homogeneous methods from GIRAFFE spectra,
with a total of 49 stars with both Na and O in NGC~6388, more in line with our
results for other GCs in our survey. 

\begin{table}
\centering
\setlength{\tabcolsep}{1mm}
\caption[]{Zn and Nd abundances for stars in Carretta et al. (2007a)}
\begin{tabular}{rccc}
\hline
Star   &   nr &  [Zn/Fe]{\sc i} &  rms \\
\hline
 77599    &   1  &   +0.323 &	\\
 83168    &   1  & $-$0.066 &  \\
101131    &   1  &   +0.476 &	\\
108176    &   1  & $-$0.283 &  \\
108895    &   1  &   +0.173 &	\\
110677    &   1  &   +0.041 &	\\
111408    &   1  &   +0.230 &	\\

\hline
\end{tabular}
\label{t:zn}
\end{table}

The interquartile range of the [O/Na] ratio, IQR[O/Na], defines the extension of
the Na-O anticorrelation and it is a good proxy for the extent of the nuclear
processing altering the composition of second generation stars (Carretta 2006,
Carretta et al. 2010). The value IQR[O/Na]=0.644 from the combined sample  of 49
stars places NGC~6388 in the region of the IQR-$M_V$ plane populated by GCs that
seems to have a too short Na-O anticorrelation when compared to their high mass
(total absolute magnitude): NGC~6441, M~15 and 47~Tuc (see Fig.~\ref{mviqr6388}.
However, when the modulation to the mass-IQR relation given by the cluster
concentration $c$ is also taken into account (as suggested in Carretta et al.
2014, see their Figs. 7 and 14), the scatter is much reduced and all these
outliers lie on the same linear relation, including NGC~6388.

\begin{figure}
\centering
\includegraphics[scale=0.40]{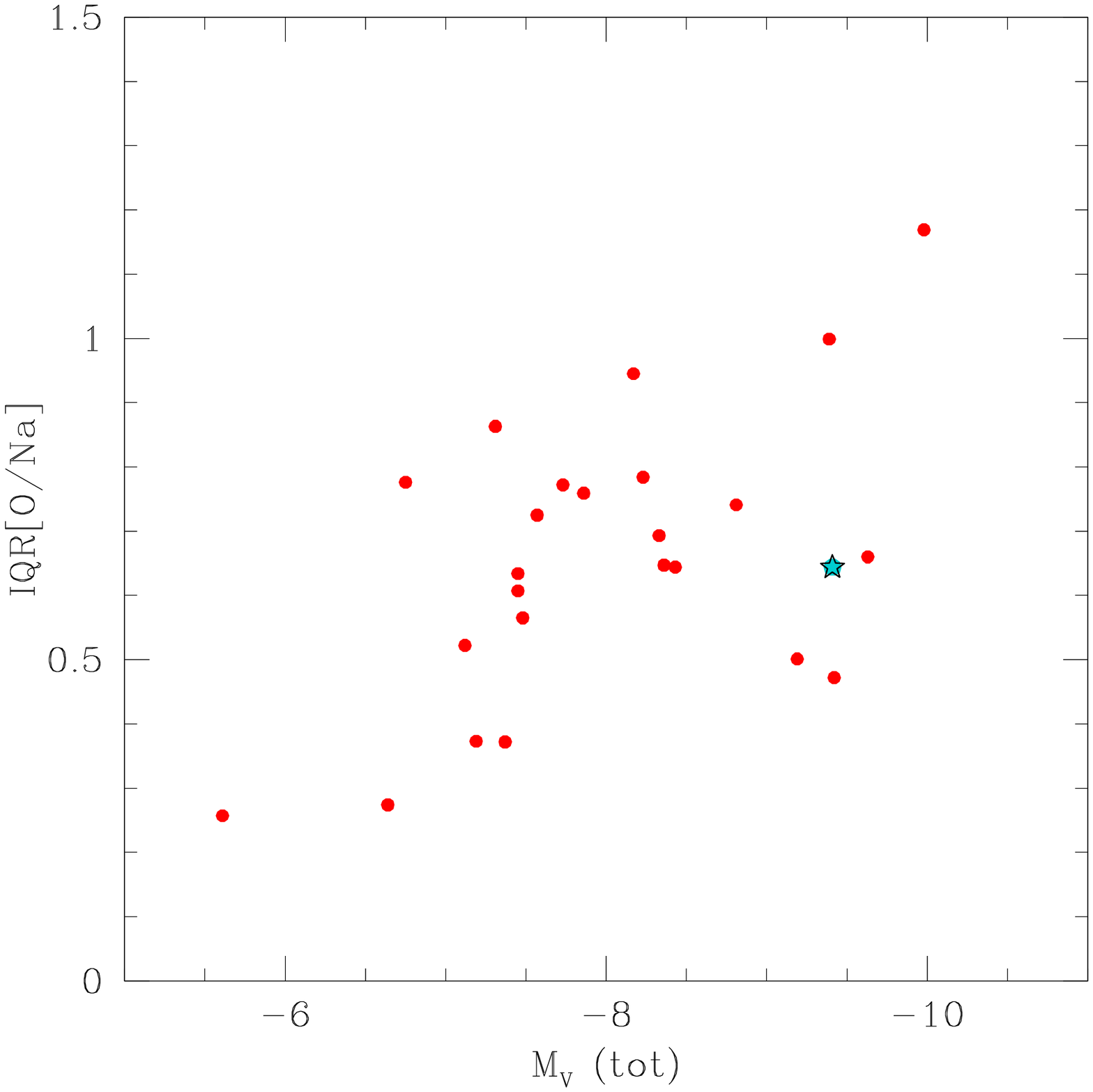}
\caption{Interquartile range of the [O/Na] ratio versus total
absolute magnitude for GCs in our FLAMES survey, with the updated value derived
in the present work for NGC~6388 (cyan star simbol).}
\label{f:mviqr6388}
\end{figure}

According to the definition given in Carretta et al. (2009b), we confirm that
NGC~6388 hosts a large population of stars with extremely modified composition
(E fraction = $20 \pm 6\%$, comparable to that of NGC~2808), whereas in about a
third of stars ($31 \pm 8\%$) the primordial composition of typical halo stars
is still preserved. The remaining stars (a fraction $49 \pm 10\%$) show an
intermediate chemical pattern.

The variation of Mg is not large. The anticorrelation of Mg with Si would
indicate a leakage from the Mg-Al cycle on $^{28}$Si (Karakas and Lattanzio
2003) which  becomes dominant only when the interior temperature T$_6 \gtrsim 65$
K (Arnould et al. 1999). Apparently, the significance of this anticorrelation in
NGC~6388 only rests on the star with the highest [Mg/Fe] and does not seem mirrored
by the expected Si-Al correlation (see Fig.~\ref{f:checku63al2}). 

\begin{figure}
\centering
\includegraphics[bb=107 154 447 709, clip, scale=0.52]{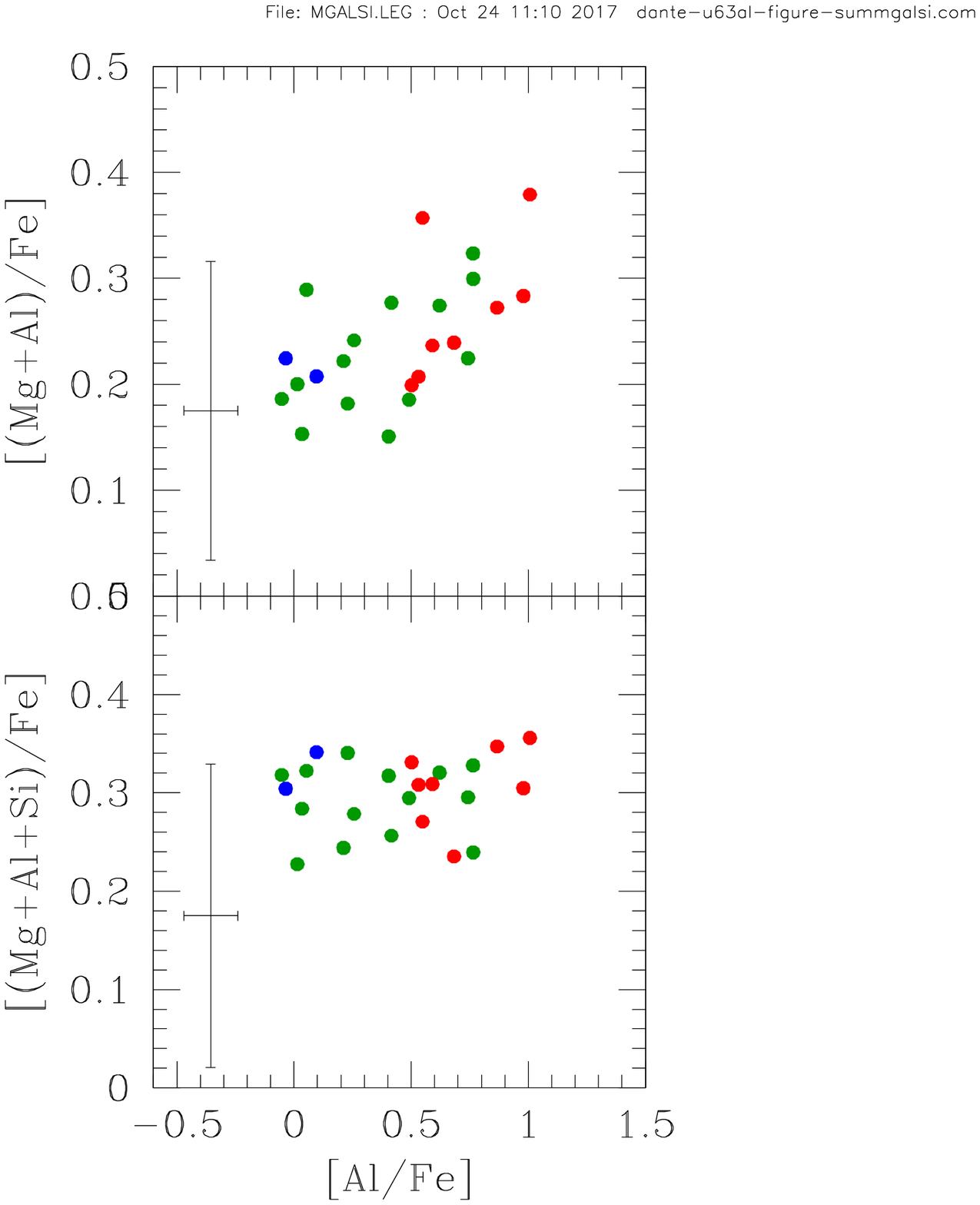}
\caption{Upper and lower panels: sum Mg+Al and Mg+Al+Si, respectively, 
as a function of the [Al/Fe] abundance. Blue, green, and red circles indicate
stars of the P, I, and E components, according to the definition by Carretta et
al. (2009b).}
\label{f:summgalsi}
\end{figure}

However, from the upper panel in Fig.~\ref{f:summgalsi} we see that as the ratio
[Al/Fe] increases, the sum Mg+Al stays approximatively constant only up to
[Al/Fe]$\simeq 0.5 - 0.6$ dex, i.e. up to typical values of the stars in the
extreme component. Afterward, the sum seems to be no more constant, and a clear
trend is visible. This trend is completely erased when also Si abundances are
taken into account, as in the lower panel in Fig.~\ref{f:summgalsi}. The ratio
[(Mg+Al+Si)/Fe] is constant (average value 0.299 dex, with $\sigma=0.037$ dex,
and indicates that, despite the absence of clear Al-Si correlation and Mg-Al
anticorrelation,  a certain amount of $^{28}$Si was produced by the  the
$^{27}$Al($p$,$\gamma$)$^{28}$Si reaction in the polluters that were active  in
FG stars of NGC~6388.

\begin{figure}
\centering
\includegraphics[scale=0.40]{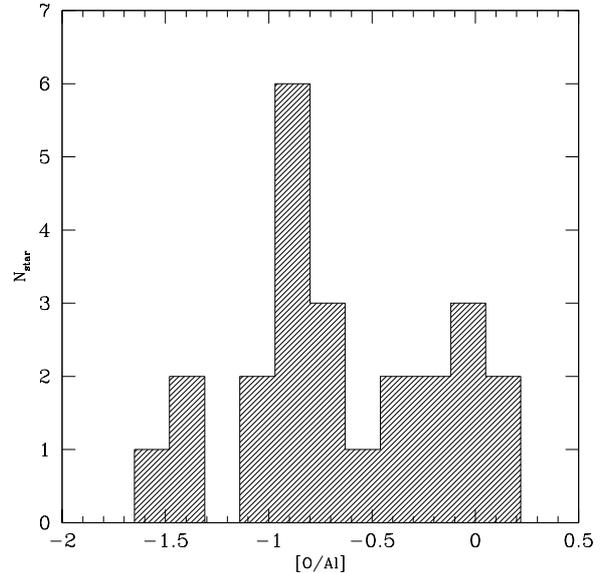}
\caption{Histogram of the distribution of stars along the Al-O anticorrelation. }
\label{f:histoal}
\end{figure}

We also observed well defined (anti)correlations between Al and O
and Na, respectively (Fig.~\ref{f:checku63al2}). Stars do not  seem uniformly
distributed in NGC~6388, in particular along the Al-O anticorrelation, as shown
in Fig.~\ref{f:histoal}. 
From the primordial to the most extreme composition the
groups are populated by 9, 12, and 3 stars, respectively, so that the PIE
fractions do match those derived from the classical definition (using O, Na),
within the associated Poisson errors.

We combined the [O/Al] ratios for the sub-samples P+I  (21 stars) and I+E (15
stars) and used the Hartigans' dip test implemented in the R package to evaluate
the unimodality/multimodality of their respective distributions. In both cases
the alternative hypothesis that the distribution is non-unimodal,
i.e. at least bimodal, is favoured.

A further confirmation is given by the Ashman's D statistics (Ashman et al.
1994), a measure of how well differentiated are two distributions. For a clean
separation, D$>2$ is required, and we obtained D=4.01 and D=4.21 for the
combination I+E and I+P, respectively. The same test applied to the PIE
distribution in the Na-O plane returns values of D=3.13 and D=3.42 (I+E and I+P,
respectively) when the combined sample GIRAFFE+UVES is used, 
and D=2.79 and D=2.87, if only the sample with UVES spectra is adopted.

In other words, this is equivalent to say that the separation between the three
groups is much larger than the dispersion within each component: the same
argument was recently used by Johnson et al. (2017) to decompose the multiple
populations in NGC 5986 into 4-5 discrete components.

\subsection{How many classes of polluters?}

The temperature range pointed out by the leakage from the Mg-Al cycle on 
$^{28}$Si (the constant Mg+Al+Si sum) does not help us much to 
efficiently discriminate among the putative candidate polluters, as no one
individually seems to be free from drawbacks strictly related to nucleosynthetic
considerations (see Bastian, Cabrera-Ziri \& Salaris 2015, and Prantzos, 
Charbonnel \& Iliadis 2017). However, our data may
at least help to constrain {\em how many} classes of polluters are required to
reproduce the observed anticorrelations of proton-capture elements in NGC~6388.

\begin{figure*}
\centering
\includegraphics[scale=0.50]{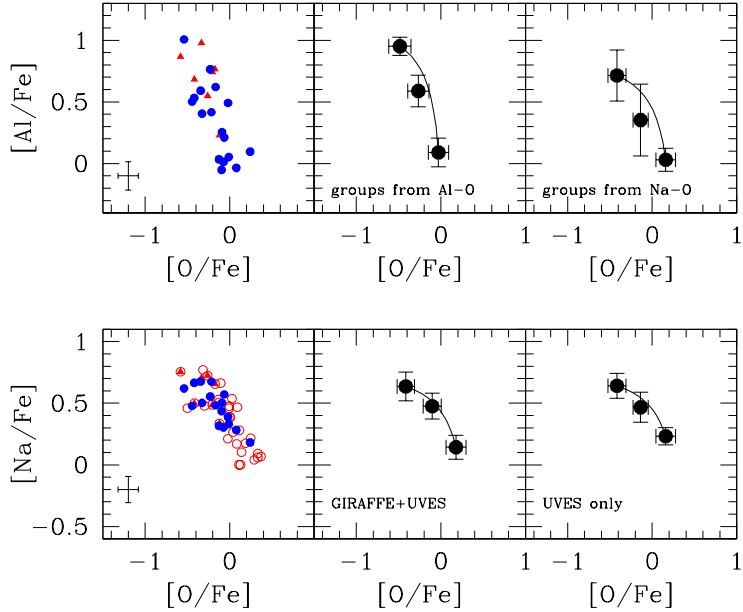}
\caption{Simple dilution models for the Al-O and Na-O anticorrelations. The
upper left panel shows the observed Al-O data. In the upper middle panel a dilution
model anchored to the mean values of the P and E components (large black points)
is shown. Error bars here are $rms$ scatter values and the separation into the P,I,
and E components is made according to the Al-O data. Upper right panel: the same, but
following the classical P, I, E classification from the Na-O anticorrelation.
In the lower panels we show the same for the Na-O data, with the middle panel referring to the
combined GIRAFFE+UVES sample, and the right panel to the UVES data only.}
\label{f:provadil}
\end{figure*}

In the middle and right panels of Fig.~\ref{f:provadil} we show simple  dilution
models (see e.g. Carretta et al. 2009b) obtained by mixing the composition of
the E component with different amounts of pristine gas, i.e. with the same
composition of stars in the primordial P component. 
By construction, the models are then anchored to the average values of the
stellar populations (large black filled circles: Al-O in the upper panels and
Na-O in the lower panels.). For Al-O we show the results using the division in
P, I, and E components derived both from the Al-O anticorrelation, where the
separation into discrete groups is more easily seen, and from the Na-O
anticorrelation. The latter is the one usually used to separate these components
following the classical definition in Carretta et al. (2009a). For the Na-O
anticorrelation we used both the combined GIRAFFE+UVES sample and the UVES data
only, as derived in the present work.

In the Na-O plane, the model passes rather well through each average value, within 
the associated $rms$ scatters. This would indicate that the intermediate I
component may be obtained by mixing the pure ejecta from a single class of
polluters (group E) with primordial gas with the same composition of the P
group. Had we used all the set of stars with GIRAFFE and UVES spectra (49
objects) the dilution model would have reproduced even better the I group.
However, this enlarged sample would not be entirely consistent with the upper
panels, where only the 24 stars from UVES spectra are available for this
exercise. 

When also Al is considered, as in the upper panels, the I component  seems to be
only marginally consistent with a single dilution model when the classification
from Al, O is used (middle panel) and it seems even less adequate if the usual
classification from the Na,O combination is adopted.

We may better quantify this finding using the same approach followed by Carretta
et al. (2012) for NGC~6752. Let's a population with intermediate composition be
obtained by mixing a fraction $dil$ of matter with E-like composition together 
with a fraction (1-$dil$) of pristine gas (P composition). Then, in the case of
a single class of polluters the value of $dil$ must be the same for all
elements:
 
\begin{equation}
dil = { {[A({\rm Mg})_I-A({\rm Mg})_P]} \over {[A({\rm Mg})_E-A({\rm Mg})_P]} } = { {[A({\rm O})_I-A({\rm O})_P]} \over {[A({\rm O})_E-A({\rm O})_P]} }
\end{equation}

where $A(el)$ are the abundances in number of atoms.

\begin{table*}
\centering
\caption[]{Dilution fractions from the GIRAFFE and UVES samples}
\begin{tabular}{lccccc}
\hline\hline
Sample      & O & Na & Mg & Al & Si \\
            & dil  & dil   & dil   & dil   & dil   \\      
\hline        
GIRAFFE+UVES & $0.64\pm0.15$ & $0.60\pm0.11$ & $0.56\pm0.34$ &  &  \\
\hline
UVES only   &  $0.67\pm0.24$ & $0.48\pm0.16$ & $0.79\pm2.29$ & $0.37\pm0.15$ & $0.91\pm0.63$ \\
groups Na-O &                &               &               &               &               \\
\hline
UVES only   &  $0.66\pm0.25$ & $0.48\pm0.14$ & $0.08\pm1.17$ & $0.34\pm0.07$ &               \\
groups Al-O &                &               &               &               &               \\
\hline
\end{tabular}
\label{t:dilu63al}
\end{table*}

\begin{figure}
\centering
\includegraphics[bb=23 157 559 707, clip, scale=0.45]{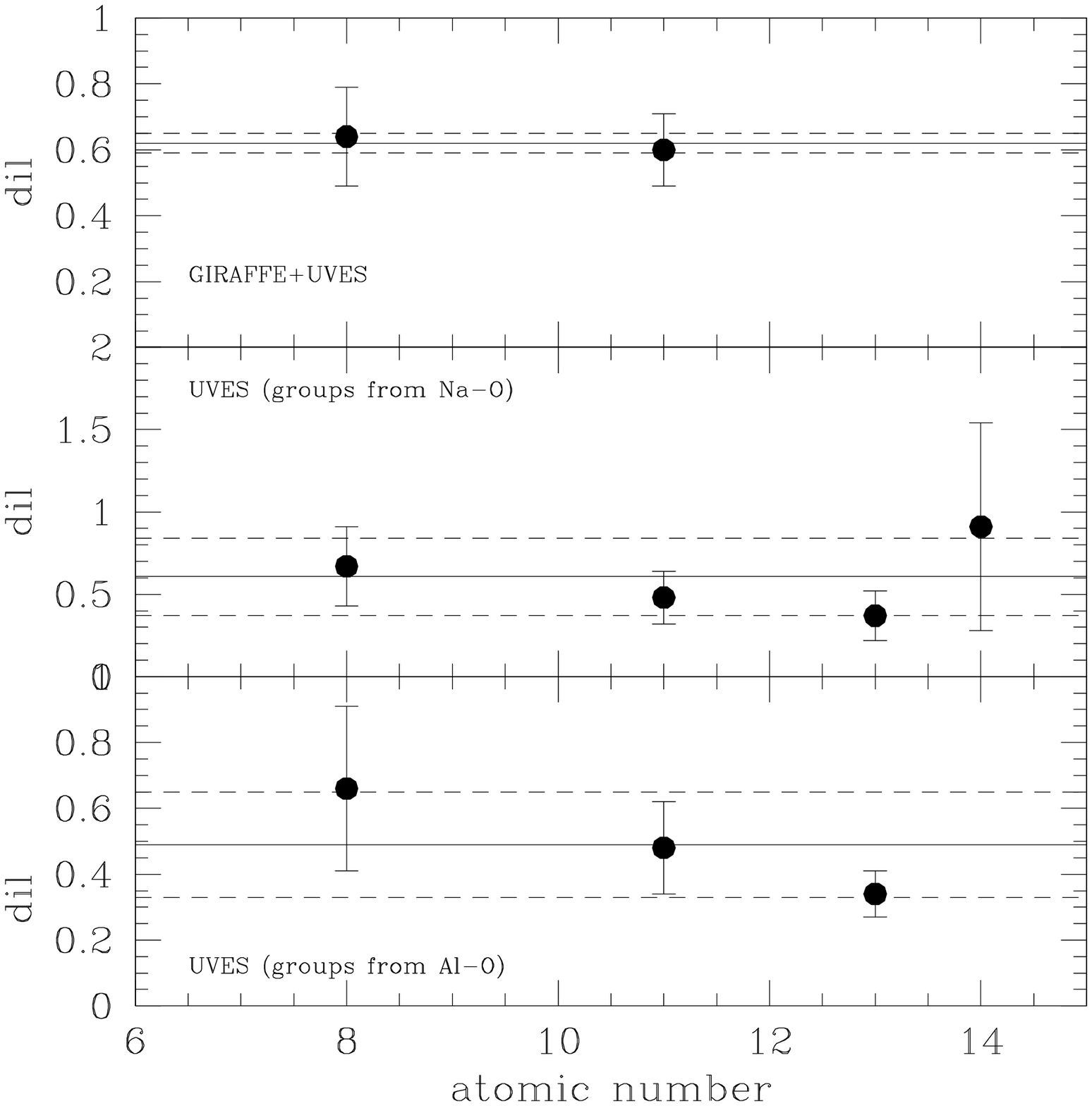}
\caption{Dilution values for different species (indicated by the atomic 
number). The reference samples are labelled in each panel. Solid and dashed
lines represent the average value of $dil$ and $\pm 1\sigma$, respectively.}
\label{f:summarydil}
\end{figure}

The results from different sample combinations and species are summarised in
Table~\ref{t:dilu63al} and in Fig.~\ref{f:summarydil}, where we plot the $dil$
values as a function of the atomic number of the considered element. Average
values from all species are represented by a solid line, together with 
$\pm 1\sigma$ values. 

What kind of conclusions can we draw from this exercise? First, we note that for
Na and O the $dil$ value seems to be the same in all cases, within the
uncertainties, although when the UVES sample is split according to the Al-O
anticorrelation the similarity is marginal (see also the middle upper panel in
Fig.~\ref{f:provadil}). However, these elements are both probes of polluters
with a moderate inner temperature ($> 40 \times 10^6$ K, ON and NeNa cycles),
and so moderate masses. 

As an indicator of higher mass polluters
we can use the MgAl cycle occurring at higher temperatures ($> 70 \times 10^6$
K, e.g. Denisenkov and Denisenkova 1989, Langer et al. 1993). Values of $dil$
for Mg are not plotted in Fig.~\ref{f:summarydil} because of the large
associated errors (see Tab.~\ref{t:dilu63al}). They are due both to the limited
size of the sample and to the small range of variation of Mg (see e.g.
Fig.~\ref{f:checku63al2}). However, the other element involved in this cycle,
Al, provides the indication that the composition of the I group is still (barely)
compatible with a single class of polluters. Taking into account
also the contribution of Si (Fig.~\ref{f:summgalsi} and middle panel of
Fig.~\ref{f:summarydil}) there is a substantial hint that more than a single
class of polluters may be required to reproduce the chemical pattern of the
stellar population with intermediate composition in NGC~6388.

Note that the largest available sample (GIRAFFE+UVES) only has Na and O
derived for proton-capture elements, and lacks Al abundances. On the other hand,
for stars with derived Al abundances, only two stars are in the P group
(separation with Na,O) and only three in the E group (separation with Al, O) and
this may make difficult to obtain clearcut results.
 A larger sample of stars with Al abundances determined together with O, Na, Mg
would be highly desirable.

\subsection{Radial distribution}

The main scenarios for the origin of multiple populations in GCs predict that 
SG stars are expected to be formed more radially concentrated, as a consequence
either of a cooling flow collecting gas at the cluster centre (e.g. D'Ercole et
al. 2010) or of the gas from equatorial discs of rotating massive stars gathering
in the cluster central regions (Decressin et al. 2010). Both photometric and
spectroscopic observations generally support this prediction, although counter-examples do
exist (up to date references may be found in Johnson et al. 2017).

A further complication is that dynamical simulations like those by Vesperini et
al. (2013) show that spatial mixing may onset when a considerable fraction
(90\% or more) of the cluster mass is lost. However, the memory of the initial
conditions may be saved, even if the half mass relaxation time is rather short 
for NGC~6388 (0.79 Gyr, Harris 1996, on line edition 2010).

\begin{figure}
\centering
\includegraphics[scale=0.40]{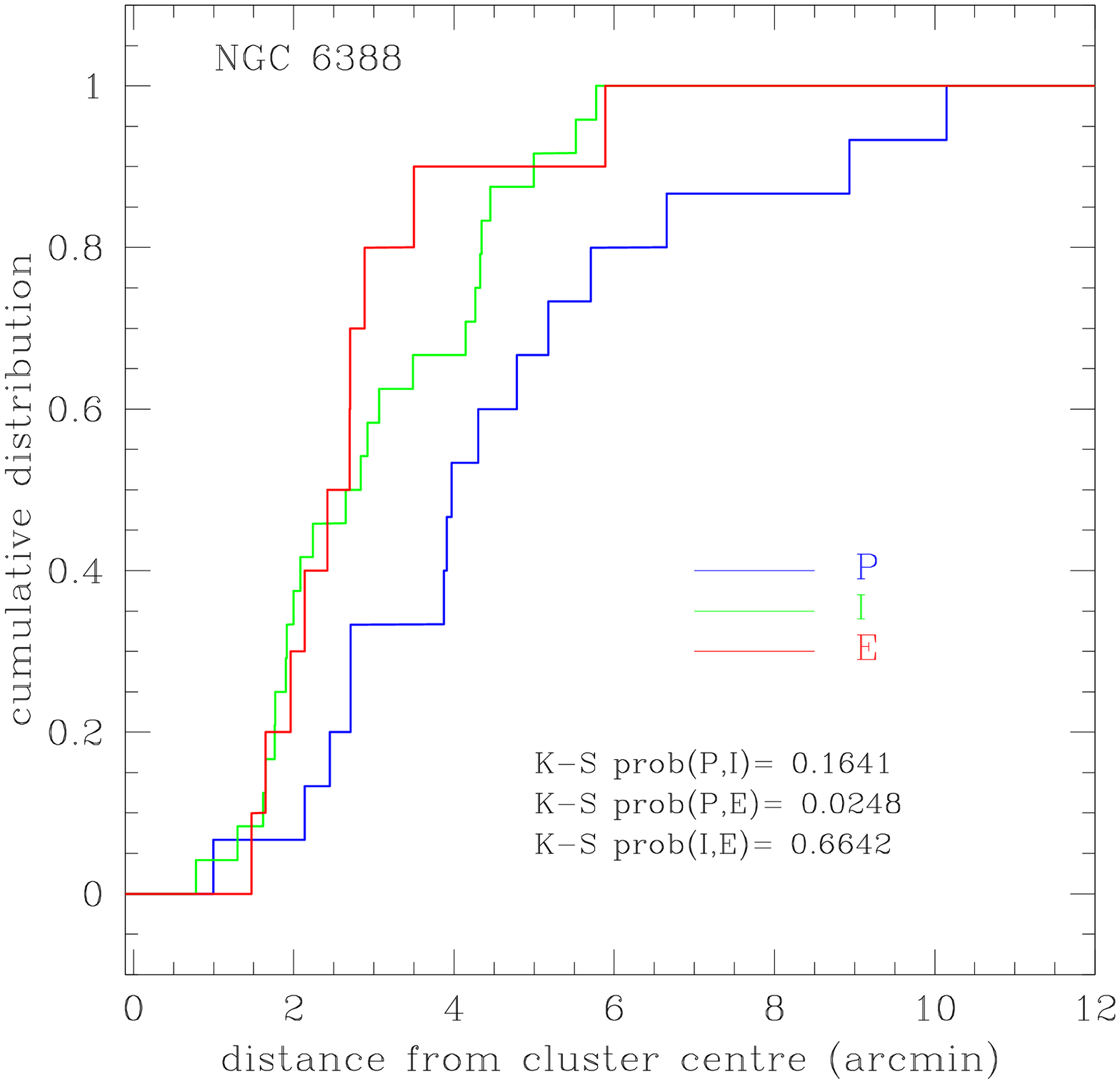}
\caption{Cumulative radial distribution of first and second generation 
stars in NGC~6388. The results of the Kolmogorov-Smirnov test between the
distribution of FG and SG stars are labelled.}
\label{f:distrPIEu63al}
\end{figure}

The comparison between the cumulative radial distribution of FG (P composition)
and SG stars (either with I or E composition) is shown in 
Fig.~\ref{f:distrPIEu63al} using the usual classification from Na, O. 
The SG stars are  more concentrated than FG stars; however, only for the
groups P and E a Kolmogorov-Smirnov test provides evidence to safely
reject the null hypothesis that the two components are extracted from the same
parent population. The comparison between P and I, and I and E, on the other hand, does not
return a statistically significant evidence.

Unfortunately, the presently available data cannot tell us more. Vesperini et
al. (2013) also showed that the radial coverage of the samples is crucial, and
they indicate that only in the cluster region around 1-2 half mass radii the
local ratio SG/FG is equivalent to the global value. In our sample we only have
one star of the P group and one of the I component within 1 arcmin from the
centre, i.e. about two times the half mass radius ($r_h=0.52$ arcmin) of
NGC~6388. The present sample is not well tailored to properly address this issue
and unfortunately the radial distribution of multiple stellar populations in
NGC~6388 has not been studied using photometry yet. We hope it will be done,
combining the central regions best observed with the HST and the external
regions well accessible from the ground.

\section{Other elements}

Although the main aim of the present work is expanding the sample useful to
study multiple stellar populations in NGC~6388, looking at the set of other
elements derived in the present analysis may be interesting. To better put this
cluster in the context of Galactic populations we compared its proton- and
$\alpha-$capture elements, as well as those of the Fe-group, with different
samples with metallicity close to that of NGC~6388.  The adopted comparison
samples are listed in Table~\ref{t:samples}, along with the symbols used in the
respective figures. All datasets are shifted to our system of solar reference
abundances using values quoted in the original papers or at the site
{\em https://github.com/NuGrid/NuPyCEE/tree/master/stellab\_data}. However, other
offsets due
to the different approaches (scales of adopted atmospheric parameters, line
lists, model atmospheres, etc) may obviously be present.

\begin{table*}
\centering
\caption[]{Samples for comparison of elements in NGC~6388 and in field stars}
\begin{tabular}{llll}
\hline
\hline
Paper                     & Sample                 & elements                                       & symbols \\
\hline

Gratton et al. (2003)     & halo+thick \& thin discs  & O, Na, Mg, Si, Ca, Ti~{\sc i}, Ti~{\sc ii}, Sc~{\sc ii}, Mn, Ni, Zn & orange filled triangles \\
Neves et al. (2009)       & thin \& thick discs       & Na, Mg, Al, Si, Ca, Ti~{\sc i}, Ti~{\sc ii}, Sc~{\sc ii}, Mn, Ni    & empty grey triangles    \\
Bensby et al. (2014)      & halo+thick \& thin discs  & O                                              & empty green triangles   \\
Bensby et al. (2005)      & thin \& thick discs	   & Zn                                             & empty green triangles   \\
Alves-Brito et al. (2010) & bulge       	   & O, Na, Mg, Si, Ca, Ti~{\sc i}                        & filled black squares    \\
Barbuy et al. (2013)      & bulge		   & Mn                                             & filled black squares    \\
Bensby et al. (2017)      & bulge		   & Ni, Zn                                         & filled black squares    \\

\hline
\hline
\end{tabular}
\label{t:samples}
\end{table*}

\begin{figure}
\centering
\includegraphics[scale=0.40]{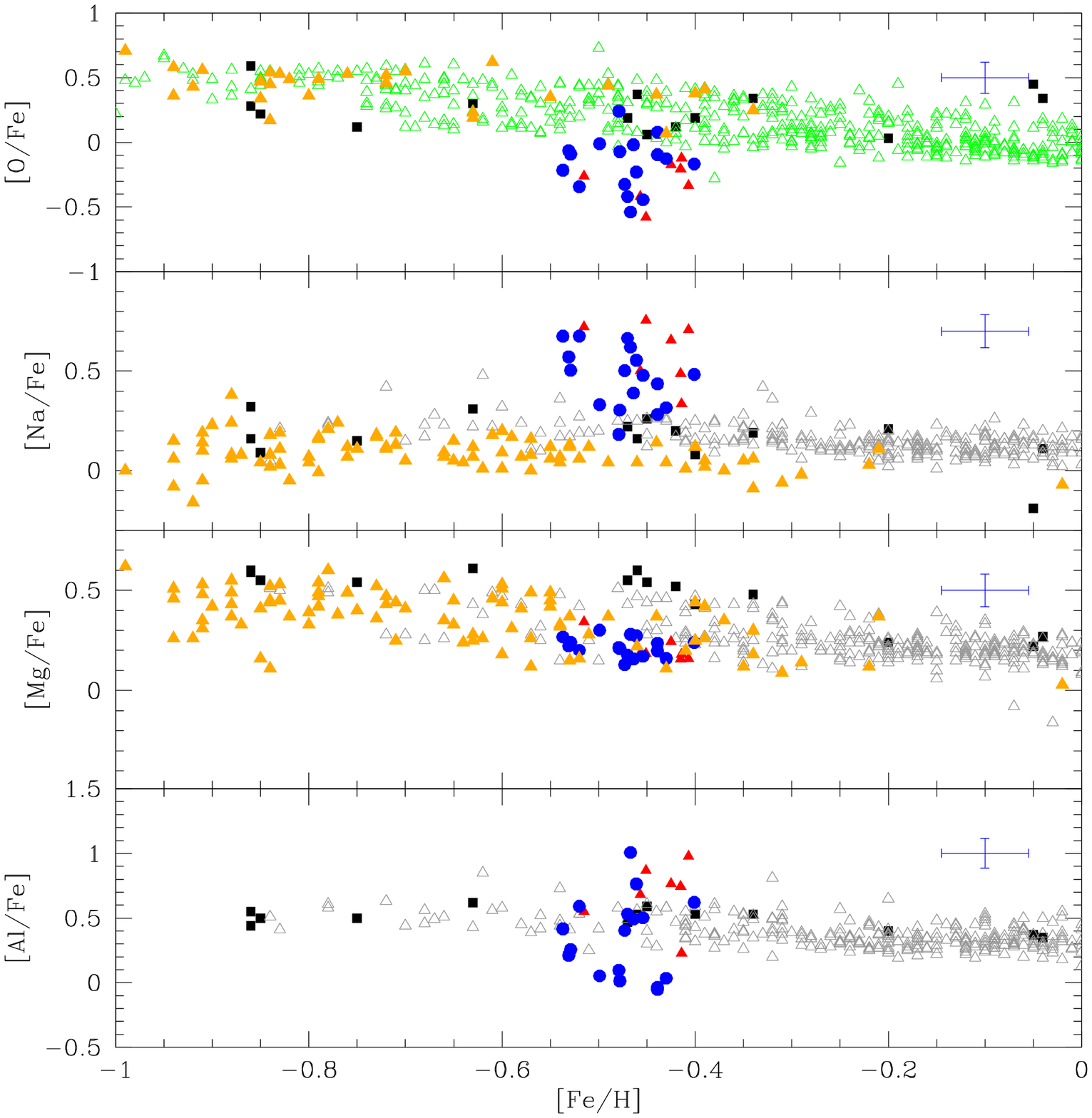}
\caption{Comparison of abundances of O, Na, Mg, and Al in NGC~6388 with
different samples of field stars. Blue circles and red filled triangles are
stars in NGC~6388. Samples, references and symbols for field stars are in
Table~\ref{t:samples}. The internal error bars displayed refer to the sample
analyzed in the present work.}
\label{f:onamgalFE}
\end{figure}

In Fig.~\ref{f:onamgalFE} one can fully appreciate the chemical signature of
a globular cluster from the clear departure of the proton-capture elements O,
Na, Mg, and Al from the patterm defined by the field stellar populations.
The O depletion and Na enhancement clearly set stars of NGC~6388 apart of any Galactic
component at [Fe/H]$\sim -0.5$ dex, either disc or bulge stars. The average
abundance of Mg in NGC~6388 is at the lower edge of any distribution, being much
lower than the typical level of bulge stars.

\begin{figure}
\centering
\includegraphics[scale=0.40]{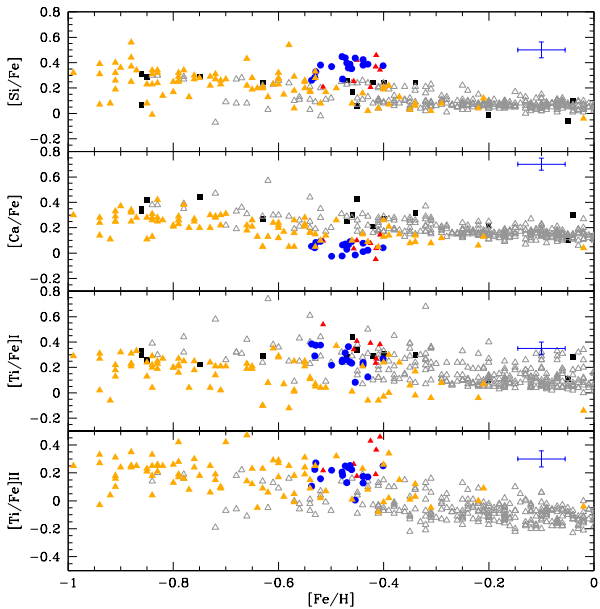}
\caption{As in Fig.~\ref{f:onamgalFE} for Si, Ca, Ti~{\sc i} and 
Ti~{\sc ii}.}
\label{f:sicatitiFE}
\end{figure}

Regarding other $\alpha-$elements (Fig.~\ref{f:sicatitiFE}), we see that the level of [Si/Fe] is higher
than in field stars, confirming the relevance of the contribution from high
temperature H-burning to the chemical pattern of stars in NGC~6388. The [Ca/Fe]
values are slightly lower, on average, than for field disc and bulge stars. The 
reasonably good agreement of the average abundances of Ti from neutral and
ionized transitions, within the errors, supports the reliability of the adopted
atmospheric parameters. They are also similar to those of field stars even if,  unfortunately,
we did not find samples of bulge stars for comparison to 
 [Ti/Fe]~{\sc ii} 
 (or [Sc/Fe]~{\sc ii}, see next figure). 

\begin{figure}
\centering
\includegraphics[scale=0.40]{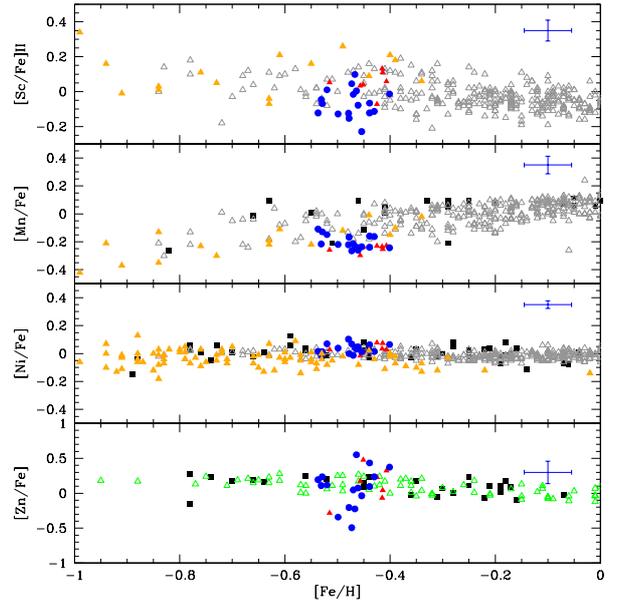}
\caption{As in Fig.~\ref{f:onamgalFE} for Sc~{\sc ii}, Mn, Ni, Zn.}
\label{f:scmnniznFE}
\end{figure}

The scatter of Sc seems to be comparable to that of field stars 
(Fig.~\ref{f:scmnniznFE}). The [Mn/Fe] ratios are more in agreement with the 
lower values found in disc stars than in bulge stars, while Ni is
indistinguishable among the various stellar populations. The Zn abundances show
a large scatter, but they rest only on a single line (see 
Table~\ref{t:fepeaku63al}). Anyway, the mean Zn abundance in NGC~6388 seems to be
in agreement with the mean trend in field stars.

We conclude reminding that NGC~6388 was found by Casetti-Dinescu et al.
(2010) to be part of a kinematic system hotter than the thick disc, and centred
on the Galactic bulge. Furthermore,  from similarities with the twin globular
cluster NGC~6441, they considered the possibility that both GCs could have
been formed
into the same dwarf galaxy, later accreted by the Milky Way. Our data and the
differences with respect to the bulge samples, could maybe be interpreted in
this framework, although more homogeneous analysis would be required concerning
field stars. 

\section{Summary and conclusions}

We analysed high resolution UVES spectra of 17 RGB member stars of the massive
and metal-rich globular cluster NGC~6388. These stars were observed mainly to
derive radial velocities, but they can be also used to derive abundances of
several elements. We obtained abundances of proton-capture (O, Na, Mg, Al, Si)
elements, $\alpha-$elements (Ca, Ti), and iron-group elements (Sc, V, Cr, Mn,
Co, Ni, Zn) for the present new sample, to be added to the similar analysis
of other 7 giants in Carretta et al. (2007a), for a total of 24 stars
with homogeneous abundances from high resolution spectra. 

We confirm that this cluster hosts one of the largest fraction (20\%) of stars 
with extremely modified composition, according to the usual classification by
Carretta et al. (2009b). About a third of stars (31\%) still have the primordial
composition of FG stars: this value is typical of what found for other GCs in
our FLAMES survey (e.g. Carretta et al. 2009a,b). The other half of stars shows
an intermediate composition.

The distribution of stars in the Al-O plane clearly presents three groups.
Statistical tests strongly support the idea that these are discrete components,
as also confirmed by the star distribution along the Na-O anticorrelation.

Thorough examination of simple dilution models for all the available light 
elements provides substantial hints that more than a single class of FG polluters
is required to account for the chemical composition of the intermediate
component. Already by simply tripling the sample of stars with all proton-capture 
elements derived homogeneously we highlighted that probably two kind of polluters of
different masses were at work: one for ordinary Na-O processing, at moderate
temperature,  the other reaching higher temperatures necessary to generate the Mg 
depletion and even some leakage on Si.
This conclusion is strongly supported by the sum Mg+Al stopping to be constant
for the stars in the extreme SG component. For this E fraction, it is
the sum Mg+Al+Si to be constant.

A more clear-cut solution of this issue is hampered by the lack of a  larger
sample of stars where the full set of the key
elements characterizing multiple populations in GC is  simultaneously available. 
This can be obtained 
even using intermediate resolution GIRAFFE spectra;  NGC~6388
is quite metal-rich and precise enough abundances of
Na, O, Mg and Al may be obtained from GIRAFFE spectra with the HR13 and HR21 setups,
respectively. 
We were granted time at ESO VLT-UT2 to observe about 60-70 secure member stars.
When all proton-capture elements will be derived for all stars from proprietary
and archival spectra, it will be possible to apply a statistical cluster analysis
to confirm or reject the existence of discrete populations and their
composition, with more precise insights on the nature and number of classes
of FG polluters in this cluster.

\begin{acknowledgements}
We wish to thank the anonymous referee for constructive suggestions that helped
to improve the paper.
We gratefully acknowledge the use of the ESO Science Archive Facility.
This publication makes use of data products from the Two Micron All Sky Survey,
which is a joint project of the University of Massachusetts and the Infrared
Processing and Analysis Center/California Institute of Technology, funded by the
National Aeronautics and Space Administration and the National Science
Foundation. This research has made use of the SIMBAD database (in particular 
Vizier), operated at CDS, Strasbourg, France, of the NASA's Astrophysical Data
System, of {\sc TOPCAT} (http://www.starlink.ac.uk/topcat/), and of  R: A
language and environment for statistical computing. R Foundation for Statistical
Computing, Vienna, Austria. ISBN 3-900051-07-0, URL http://www.R-project.org.
\end{acknowledgements}

\end{document}